\documentclass{aa}  

\usepackage{graphicx}
\usepackage{txfonts}
\usepackage{dcolumn}
\usepackage{bm}
\usepackage{xcolor}
\usepackage{acronym}
\usepackage[normalem]{ulem}
\usepackage{amsmath}

\usepackage{natbib}
\bibpunct{(}{)}{;}{a}{}{,} 

\newacro{NSE}{nuclear statistical equilibrium}
\newacro{EOS}{equation of state}
\newacro{BNS}{binary neutron star}
\newacro{NS}{neutron star}
\newacro{BH}{Black Hole}
\newacro{CC}{charged-current}
\newacro{CCSN}{core-collapse supernova}
\newacro{GW}{gravitational wave}
\newacro{dof}{degrees of freedom}

\newcommand{\refapp}[1]{Appendix~\ref{#1}}
\newcommand{\refsec}[1]{Sect.~\ref{#1}}
\newcommand{\reftab}[1]{Table~\ref{#1}}
\newcommand{\refeq}[1]{Eq.~(\ref{#1})}
\newcommand{\reffig}[1]{Fig.~\ref{#1}}

\begin{document} 

\title{Muons in the aftermath of neutron star mergers\\ and their impact on trapped neutrinos}

   \author{Eleonora Loffredo
          \inst{1,2,3}
          \and
          Albino Perego\inst{4,5}
          \and
          Domenico Logoteta\inst{6,7}
          \and
          Marica Branchesi\inst{1,2,3}
          }

   \institute{Gran Sasso Science Institute, Viale Francesco Crispi 7, 67100 L'Aquila, Italy
         \and
             INFN - Laboratori Nazionali del Gran Sasso, I-67100, L'Aquila (AQ), Italy
         \and
             INAF - Osservatorio Astronomico d'Abruzzo, Via M. Maggini snc, I-64100, Teramo, Italy
         \and
             Dipartimento di Fisica, Università di Trento, Via Sommarive 14, 38123 Trento, Italy          
         \and
             INFN-TIFPA, Trento Institute for Fundamental Physics and Applications, Via Sommarive 14, I-38123 Trento, Italy
          \and
             Dipartimento di Fisica, Università di Pisa, Largo B.  Pontecorvo, 3 I-56127 Pisa, Italy
           \and
             INFN, Sezione di Pisa, Largo B. Pontecorvo, 3 I-56127 Pisa, Italy
             }
 
  \abstract
   {In the upcoming years, present and next-generation gravitational wave observatories will detect a larger number of binary neutron star (BNS) mergers with increasing accuracy. In this context, improving BNS merger numerical simulations is crucial to correctly interpret the data and constrain the equation of state (EOS) of neutron stars (NSs).}
   {State-of-the-art simulations of BNS mergers do not include muons. However, muons are known to be relevant in the microphysics of cold NSs and are expected to have a significant role in mergers, where the typical thermodynamic conditions favour their production. Our work is aimed at investigating the impact of muons on the merger remnant.}
   {We post-process the outcome of four numerical relativity simulations of BNS mergers performed with three different baryonic EOSs and two mass ratios considering the first $15$ milliseconds after merger. We compute the abundance of muons in the remnant and analyse how muons affect the trapped neutrino component and the fluid pressure.}
   {We find that depending on the baryonic EOS, the net fraction of muons is between $30 \%$ and $70 \%$ the net fraction of electrons. Muons change the flavour hierarchy of trapped (anti-)neutrinos such that deep inside the remnant, muon anti-neutrinos are the most abundant, followed by electron anti-neutrinos. Finally, muons and trapped neutrinos modify the neutron-to-proton ratio, affecting the remnant pressure by up to $7\%$ when compared with calculations neglecting them.}
   {This work demonstrates that muons have a non-negligible effect on the outcome of BNS merger simulations, and they should be included to improve the accuracy of a simulation.}

   \maketitle
%

\section{Introduction}

The detection of the \ac{GW} event GW170817 from the inspiral phase of a \ac{BNS} merger \citep[][]{abbott17_prl} opened a new window into the microphysics of \acp{NS}. The measurement of the tidal deformability in addition to the constraint represented by the maximum measured NS mass has allowed researchers to 
put constraints on the \ac{EOS} of cold NSs \citep{abbott18_prl,De:2018uhw}. More stringent constraints have been obtained by combining information from the \ac{GW} signal with the properties of the detected electromagnetic counterparts \citep{abbott17_apjl}, specifically a short gamma-ray burst and a kilonova \citep[e.g.][]{bauswein17, margalit17, radice18}. 
The improved sensitivity planned in the next runs of observations by present GW detectors, such as the Laser Interferometer Gravitational-Wave Observatory (LIGO), the Virgo interferometer, and the Kamioka Gravitational Wave Detector (KAGRA) will increase the number of detected BNS mergers and the chance to observe their electromagnetic counterparts \citep{Abbott2020_LRR, Patricelli2022_MNRAS, Colombo2022}. The next generation of GW observatories, such as the Einstein Telescope \citep{ET} and the Cosmic Explorer \citep{Evans2021}, are expected to significantly enlarge the horizon of detectable BNS mergers and to dramatically improve the estimate of the source parameters \citep{grimm20, Maggiore2020_JCAP, Ronchini2022, Iacovelli2022}. To date, post-merger GWs from BNS mergers have not been detected, but
the Einstein Telescope and the Cosmic Explorer should reach the required sensitivity and largely increase the probability of detecting these signals. Such a discovery would strongly impact our knowledge of the nuclear EOS at high densities, even at a finite temperature \citep{sekiguchi11a, bauswein12, takami14, weih20, Perego:2021mkd, Breschi:2022PhRvL}, which is still quite uncertain. In this context, precise numerical simulations based on accurate modelling of the merger microphysics are mandatory. On the one hand, they are pivotal to interpret the collected data and to help constrain the nuclear EOS in both the zero and finite temperature regimes. On the other hand, they can provide reliable predictions of the properties of electromagnetic counterparts that can be used to optimise multi-messenger observational campaigns.

The EOS provides the relation between matter density, temperature, and the thermodynamical variables characterising a certain system. Building general-purpose EOSs for astrophysical simulations is extremely challenging because of the wide range of densities, temperatures, and charge fractions involved \citep{oertel17}. For this reason, it is important to assess the relevant \ac{dof} for a given astrophysical system. The thermodynamical conditions of BNS mergers are extreme, with temperatures reaching $T \sim 50 - 100$ MeV and densities approaching $n_b \sim 3 - 6 \ n_0$, where $n_b$ and $n_0$ are the baryon density and the nuclear saturation density, respectively \citep{sekiguchi11b, bernuzzi16, radice18b, perego19}. The dof usually included in a BNS merger simulation are nucleons, nuclei, electrons, positrons, and photons. Some simulations take into account hyperons, quarks, and/or trapped neutrinos \citep{sekiguchi11b, foucart16a, most19, radice22}. However, state-of-the-art simulations do not include muons even though the typical merger temperatures and densities allow for their presence.

Muons are already known to play a relevant role in the thermodynamics of proto-neutron stars \citep{prakash97} and have a non-negligible abundance inside cold NSs in neutrinoless $\beta$-equilibrium \citep{cohen70, cameron70, glendenning97, haensel00, haensel01, steiner05, alford10}. Their potential role in the context of BNS mergers was investigated by \citet{fore20} and \citet{alford21}. For example, it was found that weak reactions involving muons and negatively charged pions at nuclear densities $n_b \lesssim n_0$ and temperatures $T \sim 30$ MeV significantly reduce the mean free path of low energy muon (anti-)neutrinos. These processes are expected to modify the energy transport inside the merger remnant and to affect its stability \citep{fore20}. Moreover, muonic Urca processes and (anti-)neutrino scattering off muons produce a relevant contribution to the equilibration processes taking place in the post-merger remnant \citep{alford21}. Muons have already been included in some \ac{CCSN} simulations \citep{bollig17, bollig20, guo20, fischer20}. The appearance of a significant muonic fraction has been demonstrated after core bounce, and, albeit to a lesser extent, even prior to it \citep{fischer20}. 
In this context, muons induce a softening of the EOS, prompt a burst of muon neutrinos soon after the core bounce, and possibly facilitate the supernova explosion by enhancing the neutrino emission and changing the emitted neutrino spectra \citep{bollig17, guo20, fischer20}.

In addition to the nuclear EOS and the related dof, other variables, such as thermal pressure, differential rotation, and rotation rate, influence the evolution of the merger remnant \citep{Kaplan2014ApJ}. In particular, thermal pressure affects the angular velocity threshold of mass shedding and the remnant evolution during the phase of secular instability, with non-trivial implications for the time of collapse \citep{Kaplan2014ApJ}. Another important contribution to the merger thermodynamics comes from trapped neutrinos, which are present deep inside the merger remnant.
Indeed, they are relevant, especially for the implications on the stability of differentially rotating remnants. Their inclusion in BNS merger simulations is extremely challenging and requires a proper radiation transport scheme. To date, only a few BNS simulations have explicitly included them \citep[e.g.][]{foucart16a, radice22}. These simulations showed electron anti-neutrinos are the dominant trapped neutrinos and their appearance results in an increase of the electron and proton fractions.
These results are consistent with the outcome of \citet{perego19}, where the relevance of the trapped neutrino component was quantified within a post-processing approach. By changing the proton-to-neutron content, trapped neutrinos induce a pressure decrease of approximately $5\% - 10 \%$ inside the remnant \citep{perego19}, so they may speed up the collapse of massive binaries to black holes. Nevertheless,  muonic and tauonic (anti-)neutrinos have been treated at the same level as heavy-lepton (anti-)neutrinos in previous works because muons were not included in the microphysics modelling. To the best of our knowledge, there are no BNS merger simulations taking into account the coupling between neutrinos and muons.

In the present work, we develop a post-processing technique to quantify the fractions of muons and trapped neutrinos in the remnants of BNS mergers in order to assess their combined effect on fluid pressure. We apply our analysis to the outcome of four fully relativistic numerical simulations employing three different microphysical EOSs and two different mass ratios. We aim to show that the production of muons and the trapping of neutrinos modify the remnant pressure, which changes asymmetrically in space with respect to the case in which muons and neutrinos are neglected, depending on the assumed baryonic EOS and binary mass ratio.

This paper is structured as follows. In \refsec{sec:analytical_estimates}, we briefly discuss the mechanism of muon production and neutrino trapping in \ac{BNS} mergers and estimate the typical timescales and energies involved. 
In \refsec{sec:method}, we introduce our method based on a post-processing approach.
From \refsec{sec:muons_remnant} to \refsec{sec:pressure}, we present our results, analysing the abundance of muons in the remnant (\refsec{sec:muons_remnant}), the trapping of neutrinos (\refsec{sec:nu_trapping}), and how muons and trapped neutrinos affect the remnant pressure (\refsec{sec:pressure}). 
In \refsec{sec:discussion}, we discuss our results, comparing them to state-of-the-art simulations. Finally, we summarise our work and present our conclusions in \refsec{sec:conclusions}.


\section{Preliminary estimates}\label{sec:analytical_estimates}
\subsection{Muons in \ac{BNS} mergers}
The fraction of muons in a \ac{BNS} remnant is determined both by the fraction of muons in the two cold \acp{NS} before merger and by the efficiency of (anti)muon production during and after merger.
We consider as an example a fluid element taking part in a BNS merger. During the inspiral, muons are already present in the two cold \acp{NS} for sufficiently high densities when the electron chemical potential, $\mu_e$, exceeds the muon rest mass, $m_\mu c^2 \sim 106$ MeV \citep{cohen70, cameron70, glendenning97, haensel00, haensel01, steiner05, alford10}. For relativistic degenerate electrons, the former can be estimated as:
\begin{equation}
    \label{eq: electron chemical potential}
    \mu_{e^-} \sim p_{{\rm F},e} c \approx 131.5~{\rm MeV} \left(\dfrac{Y_{e^-}}{0.05}\right)^{1/3} \left(\dfrac{n_{b}}{0.2~{\rm fm^{-3}}} \right)^{1/3} \, ,
\end{equation}
where $p_{{\rm F},e}$ is the Fermi impulse of electrons, while $Y_{e^-} = n_{e^-}/n_b$ and $n_{e^-}$ are the electron fraction and the electron number density, respectively. 
Clearly, for neutrinoless $\beta$-equilibrated nuclear matter, $Y_{e^-} \sim 0.05-0.1$, muons are expected to be present as non-relativistic degenerate particles already close to saturation density. A simple estimate of their chemical potential, $\mu_{\mu^-}$, is given by:
\begin{equation}
    \label{eq: muon chemical potential}
    \mu_{\mu^{-}} \sim m_{\mu}c^2 + \frac{p_{{\rm F},\mu}^2}{2 m_{\mu}} \approx \left[ 106 + 28 \left(\dfrac{Y_{\mu^-}}{0.01}\right)^{2/3} \left( \dfrac{n_{b}}{0.2 \rm{~fm^{-3}}}\right)^{2/3} \right] {\rm MeV},
\end{equation}
where $p_{{\rm F},\mu}$ is the Fermi impulse of muons,
while $Y_{\mu^-} = n_{\mu^{-}}/n_b$  and $n_{\mu^{-}}$ are the fraction of muons and the muon number density, respectively.\\
We next investigate whether the typical thermodynamical conditions in \ac{BNS} mergers allow for the creation of muons. During the merger, nuclear matter is heated and the average photon energy, 
$E_\gamma \approx 2.7~ k_{\rm B} T $,
overcomes the muon mass threshold when the temperature reaches $k_{\rm B} T \gtrsim 40$ MeV. Therefore, in this temperature regime, thermal processes such as
\begin{equation*}
    \gamma + \gamma \rightarrow \mu^{-} + \mu^{+} \, , \qquad e^{+} + e^{-} \rightarrow \mu^{-} + \mu^{+} 
,\end{equation*}
drive the creation of $\mu^{\pm}$ pairs.
Additionally, neutrino pairs of all flavours are produced via thermal processes \citep[see e.g.][]{Haft:1993jt,ruffert96}
\begin{equation*}
    \gamma \rightarrow \nu + \bar{\nu} \qquad e^{+} + e^{-} \rightarrow \nu + \bar{\nu}~,
\end{equation*}
or nucleon-nucleon bremsstrahlung \citep[see e.g.][]{Hannestad:1997gc}
\begin{equation*}
    N + N \rightarrow N + N + \nu + \bar{\nu}~,
\end{equation*}
where $\nu$ denotes any flavour neutrino, while $N$ is a nucleon. For example, in the case of $e^{\pm}$ annihilation, the average energy of the neutrino pairs is \citep{Cooperstein:1988fz}:
\begin{equation}
    E_{\nu \bar{\nu}} = k_{\rm B} T \left[ \frac{F_4(\eta_{e^-})}{F_3(\eta_{e^-})} + \frac{F_4(- \eta_{e^-})}{F_3(- \eta_{e^-})}  \right]
,\end{equation}
where $\eta_{e^-} = \mu_{e^-}/k_{\rm B} T$ is the (relativistic) electron degeneracy parameter and $F_n(\eta)$ is the Fermi function of order $n$. For very degenerate electrons ($\eta_{e^-} \gg 1$) and accounting for \refeq{eq: electron chemical potential}
\begin{equation}
    E_{\nu \bar{\nu}} \sim k_B T \left[ \frac{4}{5} \eta_{e^-}  + 4 \right] \gtrsim m_\mu c^2 \, .
\end{equation}
Even when muonic (anti-)neutrinos thermalise, their average energy is
\begin{equation}
    E_{\nu_{\mu}} \sim \frac{F_3(0)}{F_2(0)} k_B T \approx 126~{\rm MeV} \left( \frac{T}{40 {\rm MeV}} \right) \gtrsim m_{\mu}c^2 \, .
\end{equation}
Thus, once high energy $\nu_{\mu}$ and $\bar{\nu}_\mu$ are formed, (anti)muons can be created via \ac{CC} semi-leptonic reactions
\begin{equation*}
    \nu_\mu + n \rightarrow p + \mu^- \, , \qquad \bar{\nu}_\mu + p \rightarrow n + \mu^{+} ,
\end{equation*}
as in the case of CCSNe \citep{bollig17, guo20, fischer20}. In addition, neutron decay,
$    n \rightarrow p + \mu^{-} + \bar{\nu}_\mu $,
plays a relevant role in the high density-low temperature regime \citep{alford21} typical of the remnant core. Finally, purely \ac{CC} leptonic processes, such as $ {\nu}_\mu + e^{-} \rightarrow \nu_e + \mu^-$,
and inverse muon decay can enhance the abundance of muons at low average neutrino energies $\lesssim 50$ MeV since the amount of electrons exceeds the amount of positrons \citep{guo20, fischer20, alford21}.

In general, matter in \ac{BNS} mergers is not in weak and thermal equilibrium, and it is not obvious that weak reactions are fast enough to allow for muon creation at short enough timescales everywhere inside the remnant \cite[see e.g.][]{hammond22}.
We estimate the rate at which these reactions occur by considering the reaction $\nu_\mu + n \rightarrow p + \mu^-$ as a representative example. In the zero momentum transfer limit, neglecting the momentum transferred to nucleons and assuming $|\mathbf{p}_N| \ll m_N$, where $\mathbf{p}_N$ and $m_N$ are the nucleon momentum and mass, respectively, the reaction rate of $\nu_\mu + n \rightarrow p + \mu^-$ is at leading order: 
\begin{eqnarray}
R_{\nu_\mu + n \rightarrow p + \mu^{-}} = \dfrac{G^2 c}{\pi ~ (\hbar c)^4} \left(g_V^2 + 3 g_A^2\right) \omega_{np} \left[1 - f_\mu(E_{\nu_\mu} + Q)\right]  \times \nonumber\\
 \times \left(E_{\nu_\mu} + Q \right)^2 \left[1 - \dfrac{m_\mu^2}{\left(E_{\nu_\mu} + Q \right)^2}\right]^{1/2}
,\end{eqnarray}
where $G$ is the Fermi constant, $g_V$ and $g_A$ are the weak couplings, $Q = m_n-m_p$ is the neutron-proton mass difference, $f_i(E_i)$ is the Fermi-Dirac distribution function of particle $i$ with energy $E_i$, and $\omega_{np}$ is given by
\begin{equation}
    \omega_{np} = \dfrac{n_p - n_n}{\exp\left[(\mu'_p - \mu'_n)/k_B T\right]-1}~,
\end{equation}
where $n_N$ and $\mu'_N$ indicate the number density and the non-relativistic chemical potential of nucleons $N = p,n$, respectively. An order of magnitude estimate of the reaction rate can be easily obtained in the non-degenerate regime, $\omega_{np} \rightarrow n_n$, by neglecting the neutron-proton mass difference\footnote{The difference between the neutron and the proton self-energies plays an important role at $n_b > n_0$, as it lowers the $\nu_\mu$ energy threshold for muon production. We have neglected this effect in our estimates.} and considering neutrino energies $E_{\nu_\mu} \simeq \mu_{\mu^-} \gtrsim m_\mu c^2$ (see \refeq{eq: muon chemical potential}). 
The inverse rate thus gives the muon production timescale:
\begin{equation}
       R^{-1}_{\nu_\mu + n \rightarrow p + \mu^{-}} \simeq 2 \times 10^{-9}{\rm s} \left(\dfrac{Y_n}{0.8}\right)^{-1} \left(\dfrac{n_b}{0.1~{\rm fm}^{-3}}\right)^{-1} \left(\dfrac{E_{\nu_\mu}}{120~{\rm MeV}}\right)^{-2}.
\end{equation}
We note that $R^{-1}_{\nu_\mu + n \rightarrow p + \mu^{-}} \ll t_{\rm{dyn}}$ for high enough densities and neutrino energies where $t_{\rm{dyn}} \sim 10^{-4}$s is the merger dynamical timescale \citep[see e.g.][]{Radice2020}. Because of the high degeneracy of neutrons, \ac{CC} semi-leptonic reactions are expected to favour the production of $\mu^-$ over $\mu^+$. Accordingly, the opacity of $\nu_\mu$ to \ac{CC} semi-leptonic reactions exceeds that of $\bar{\nu}_\mu$ \citep[at least in cases with high enough neutrino energy; see also the discussion in][]{guo20, fischer20}, and the net muon fraction becomes enhanced.

\subsection{Neutrino trapping}
In this section, we provide an estimate of the neutrino diffusion timescale from a \ac{BNS} merger remnant to explore the conditions in which neutrinos can be considered trapped, that is,
when their diffusion timescale, $t_{\rm{diff}}$, exceeds the dynamical timescale, as well as the conditions in which trapped neutrinos are in thermal and weak equilibrium with matter.
Since neutrino scattering off nucleons is among the largest sources of opacity for all neutrino flavours, 
we considered an approximate expression of its corresponding mean free path, $\lambda_{\nu N} (E_\nu) \sim 1/(n_b~\sigma_{\nu N}(E_\nu))$, where
\begin{equation}
\sigma_{\nu N} (E_\nu) = \dfrac{1}{4} \sigma_0 \left( \dfrac{E_\nu}{m_e c^2} \right)^2
,\end{equation}
gives the $\nu-N$ scattering cross section and $\sigma_0 \approx 1.76 \times 10^{-44} \rm{cm}^2$ \citep{shapiro83}.
Using random-walk arguments, we estimated $t_{\rm{diff}}$ as
\begin{equation}
    t_{\rm{diff}} \sim 3 \tau(E_\nu) \dfrac{{\ell}}{c} ~,
\end{equation}
where $\ell$ is the characteristic length scale of the diffusion process and $\tau$ is the optical depth, which we estimated as
$\tau \sim \ell/\lambda$. For example, in the case of thermal neutrinos diffusing from a remnant core of size $R_{\rm{NS}}$,
\begin{equation}\label{eq:diff_NS}
    t_{\rm{diff,NS}} \hspace{-0.7mm} \sim \hspace{-0.7mm} 3 \dfrac{R_{{\rm NS}}^2}{\lambda_{\nu N}(E_\nu) c} \hspace{-0.8mm} \approx \hspace{-0.8mm} 1.5~{\rm s} \left( \dfrac{n_b}{0.1~{\rm fm}^{-3}}\right) \left( \dfrac{R_{{\rm NS}}}{15~{\rm km}}\right)^2 \left(\dfrac{k_{\rm B} T_{\rm NS}}{20~\rm{MeV}}\right)^2 ,
\end{equation}
and $R_{{\rm NS}}$ and $T_{{\rm NS}}$ are the characteristic radius and temperature of the massive \ac{NS} remnant, respectively \citep[see e.g.][]{Bernuzzi2020}.
We can repeat the estimate for thermal neutrinos diffusing from the innermost part of the disc surrounding the massive central remnant:
\begin{equation}\label{eq:diff_disc}
    t_{{\rm diff,disc}} \hspace{-0.7mm} \sim \hspace{-0.7mm} 3 \dfrac{H_{{\rm disc}}^2}{\lambda_{\nu N}(E_\nu) c} \hspace{-0.8mm} \approx \hspace{-0.8mm} 20{\rm ms} \left( \dfrac{n_b}{0.01{\rm fm}^{-3}}\right) \left( \dfrac{H_{{\rm disc}}}{20{\rm km}}\right)^2 \left(\dfrac{k_{\rm B} T_{\rm disc}}{5{\rm MeV}}\right)^2
,\end{equation}
where $H_{\rm disc}$ and $T_{\rm disc}$ are the characteristic disc height and temperature, respectively.
We note that the average energy of thermal neutrinos at $k_{\rm B} T \sim 5~\rm{MeV}$ is $E_\nu \sim 15~\rm{MeV}$,
which corresponds to the typical energy of the neutrinos emitted at infinity by \ac{BNS} merger remnants
\citep[e.g.][]{Ruffert:1998, Rosswog:2003rv, Sekiguchi:2016, foucart16a, foucart16b, cusinato21}.
As implied by \refeq{eq:diff_disc}, the diffusion timescale of neutrinos with energy $E_\nu \sim 15~\rm{MeV}$ from the disc becomes comparable to the dynamical timescale around $n_b \sim 5 \times 10^{-3} \rm{fm}^{-3}$, corresponding to a rest mass density $\rho = m_{\rm{amu}} n_b \sim 10^{12} \rm{g~cm}^{-3}$, where $m_{\rm amu}$ is the atomic mass unit. Other interactions involving neutrinos in addition to $\nu-N$ scattering provide further opacity. Thus, the above estimates can be understood as conservative upper limits for the trapping density.

Due to the presence of inelastic processes (such as absorption of neutrinos on free nucleons, inverse pair processes, and scattering off electrons and positrons), neutrinos are not only efficiently trapped, but they couple to matter and photons if the corresponding mean free path is short enough.
The situation is however complicated by the fact that neutrinos have a spectral distribution and that cross sections have a non-trivial energy dependence. As a result, neutrinos of different flavours and energies decouple from matter at different locations. Rest mass density has been shown to be the most relevant matter property in determining the location of both the last scattering surface and the surfaces where weak and thermal equilibrium freezes out \citep{endrizzi20}.
In particular, according to \citet{endrizzi20}, we observe that for densities larger than $10^{12}~{\rm g~cm^{-3}}$, almost all relevant neutrinos are within the last scattering surface (Fig. 8), while for densities larger than $10^{13}~{\rm g~cm^{-3}}$, they are also in thermal and weak equilibrium (Fig. 10). In the case of $\nu_e$ and $\bar{\nu}_e$, these surfaces are even characterised by densities one order of magnitude smaller.
Thus, we defined a limiting density $\rho_{\rm lim}$ such that neutrinos can be considered as a trapped gas if $\rho \gtrsim \rho_{\rm lim}$. For electron (anti-)neutrinos, we considered $\rho_{\rm{lim}, e} = 10^{11} \text{g}~\text{cm}^{-3}$, while for muonic and tauonic (anti-)neutrinos, we set $\rho_{\rm lim,x} = 10^{12} \text{g}~\text{cm}^{-3}$.
Moreover, at $\rho \gtrsim 10 \times \rho_{\rm lim}$ neutrinos can be approximately modelled as a gas in equilibrium with other constituents of matter.

\section{Method}\label{sec:method}
We present our method to estimate the fraction of muons and their impact on the trapped neutrino component in \ac{BNS} merger remnants by post-processing the outcome of a significant sample of numerical relativity simulations. First, we review the properties of the simulations considered in this work (\refsec{sec:simulation_sample}). Next, we describe how we model the nuclear \ac{EOS}, including muons and trapped neutrinos (\refsec{sec:eos_modelling}). Finally, we explain our post-processing technique by arguing for the physical rationale and the limits of validity (\refsec{sec:post_proc}).
\subsection{The simulation sample}\label{sec:simulation_sample}
\begin{figure*}[]
    \centering
    \includegraphics[scale=0.56]{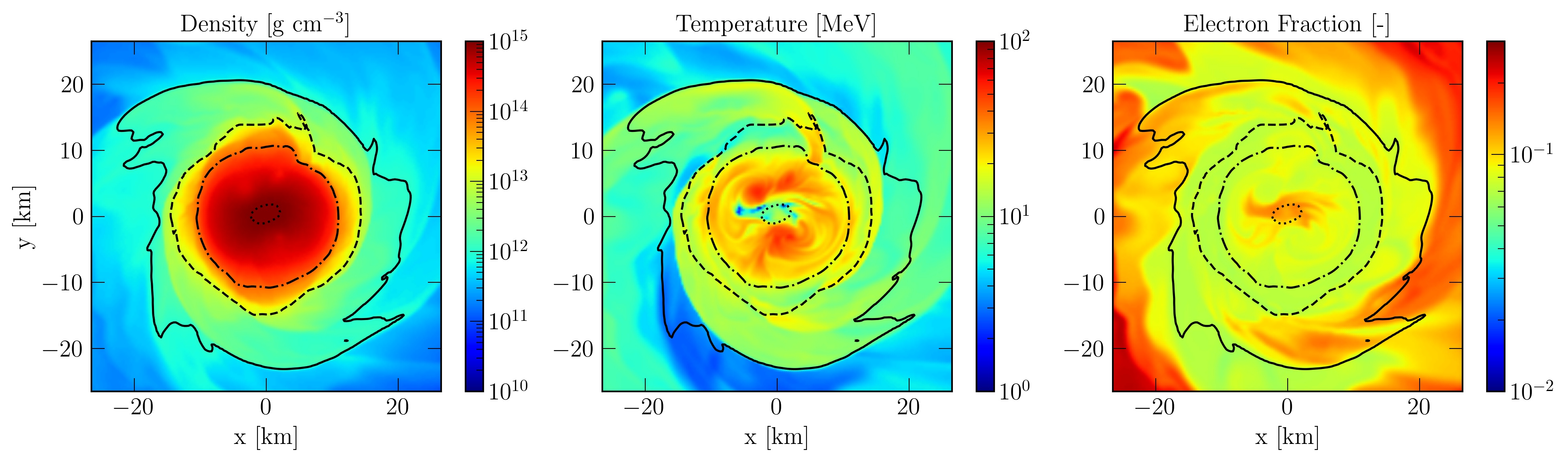}
    \caption{Outcome of the numerical simulation (BLh, 1.00) at 4.6 ms after merger. The left, central, and right panels show the rest mass density, the temperature, and the electron fraction on the equatorial plane, respectively. The black lines mark isodensity contours corresponding to $10^{12} \rm{g~cm}^{-3}$ (solid line), $10^{13} \rm{g~cm}^{-3}$ (dashed line), $10^{14} \rm{g~cm}^{-3}$ (dashed-dotted line), and $10^{15} \rm{g~cm}^{-3}$ (dotted line).}
    \label{fig:1}
\end{figure*}

We considered the outcomes of four \ac{BNS} merger simulations in numerical relativity targeted to GW170817 (see \reftab{tab:sim_sample}) published in \citet{nedora19}, \citet{bernuzzi20}, and \citet{nedora21}. The term $M_{A,B}$ refers to the gravitational mass of each of the two \acp{NS} at infinity such that the binary mass ratio is defined as $q = M_B/M_A \geq 1$.
More details about the simulations and their numerical setup can be found in the references listed in \reftab{tab:sim_sample} and in \citet{radice18b}. In the following paragraphs, we provide a summary of the simulation properties necessary to understand our post-processing procedure.

\begin{table*}[]
\caption{ List of the \ac{BNS} merger simulations considered in this work. The nuclear equation of state is listed under EOS, $M_A$ and $M_B$ are the gravitational masses of the binary, and $q = M_B/M_A$ is the binary mass ratio.}
\label{tab:sim_sample}
\centering
\begin{tabular}{lccccll}
\hline \\ [-1.5ex]
EOS & $M_A~[M_\odot]$ & $M_B~[M_\odot]$ & $q$ & Viscosity & Remnant & Refs.\\
\hline \\ [-1.5ex]
BLh & $1.364$ & $1.364$ & $1.00$ & NO & Long-lived & \cite{nedora21} \\
DD2 & $1.364$ & $1.364$ & $1.00$ & NO & Long-lived & \cite{nedora19} \\
SFHo & $1.364$ & $1.364$ & $1.00$ & YES & Very short-lived & \cite{bernuzzi20} \\
BLh & $1.146$ & $1.635$ & $1.43$ & YES & Long-lived & \cite{nedora21}\\
\hline
\end{tabular}
\end{table*}

Our simulation sample contains three symmetric ($q=1$) simulations. These simulations were performed with three different nuclear \ac{EOS}s, namely, BLh \citep{bombaci18, logoteta21}, HS (DD2; \citealt{typel10, hempel10}), and SFHo \citep{steiner13}. In the following sections, we refer to the second EOS simply as DD2. Details about these nuclear \acp{EOS} are provided in \refsec{sec:baryons}. These simulations predicted quite different merger outcomes. In particular, for SFHo the remnant collapsed into a black hole within a few milliseconds after the merger, while it survived for at least $90$ ms in the other two cases. In addition, we considered a simulation of a significantly asymmetric \ac{BNS} merger ($q=1.43$) using the BLh \ac{EOS}. 
Within this sample, we could explore how the presence of muons and trapped neutrinos correlates with the properties of the nuclear \ac{EOS} and with the mass asymmetry degree of the binary. 
Additionally, half of the simulations included the effect of physical viscosity of magnetic origin \citep{Radice:2016dwd}. The simulation sample is however too small to address the impact of viscosity on our results, and we leave such an investigation to future studies.
In the following sections, we refer to each simulation by specifying the nuclear \ac{EOS} and $q$ according to the notation: (\ac{EOS}, $q$).

All simulations modelled \ac{NS} matter as made of neutrons, protons (both free and bound in nuclei), photons, electrons, and positrons. No muons and trapped neutrinos were considered.
All the simulations included the effect of neutrino radiation through a leakage plus M0 scheme, whose details can be found in \citet{Radice:2016dwd} and \citet{radice18b}. Electron neutrinos and anti-neutrinos were considered separately, while there was no distinction between muonic and tauonic (anti-)neutrinos, which were treated as a single species (heavy-lepton neutrinos).
The leakage prescription was used to compute the net rate of change in the internal energy and in the lepton fraction for matter in optically thick conditions due to neutrino diffusion.
Even though the particle and energy density of trapped neutrinos were estimated by the leakage algorithm in order to compute the neutrino diffusion rates, they were not dynamically evolved, and their contribution to the thermodynamical state of matter was not explicitly taken into account in the simulations. This also implies that the energy density and pressure of the neutrinos were not included in the stress-energy tensor of matter and radiation.

In our analysis, we considered snapshots of the computational domain from the post-merger phase in the time interval $\sim 5 - 15~$ms post-merger where the merger time is defined by the maximum in the $\ell=m=2$ mode in the \ac{GW} waveform.
In this phase, the merger remnant is still in the \ac{GW}-dominated phase \citep[e.g.][]{bernuzzi16,Zappa2018}, but the initial, highly dynamical transients that characterise the very first milliseconds after the merger have disappeared and a disc around the central remnant has formed.  From each snapshot, we extracted the full 3D profile of the local baryon number density $n_{b}^{{\rm sim}}$, net electron fraction $Y_{e}^{{\rm sim}}$, and energy density $e_{\rm sim}$, as directly obtained by the simulation. The net electron fraction was defined as $Y_e = (n_{e^-} - n_{e^+})/n_b$, where $n_{e^-}(n_{e^+})$ is the electron (positron) number density. By using the same \ac{EOS} used in the simulation, we could reconstruct the full thermodynamic states anywhere inside the
computational domain, including the values of the temperature $T_{{\rm sim}}$, chemical potentials $\mu_j^{{\rm sim}}$, particle fractions $Y_j^{{\rm sim}}$, partial pressures $P_j^{{\rm sim}}$, and total pressure $P_{{\rm sim}}$, where the subscript $j = \{n, p, e^{\pm}, \gamma\}$ indicates neutrons, protons, electrons, positrons, and photons.
In \reffig{fig:1}, we show as an example the rest mass density $\rho_{{\rm sim}} = m_{\rm amu} n_b^{\rm sim}$, the temperature, and the electron fraction from the outcome of the simulation (BLh, 1.00) at approximately 5 ms after merger. The dense and cold core of the remnant can be recognised, characterised by $\rho_{{\rm sim}} \gtrsim 10^{15} \rm{g~cm}^{-3}$, $T_{\rm sim} < 20$ MeV, and $Y_e^{\rm sim} \sim 0.1$. The core is surrounded by warm matter in the density regime $\rho_{{\rm sim}} \sim 10^{14} - 10^{15} \rm{g~cm}^{-3}$, with the temperature reaching up $T_{\rm sim} \sim 50$ MeV in correspondence to the hot spots and a slightly smaller electron fraction $Y_e^{\rm sim} \sim 0.07 - 0.08$. At a lower density $\rho_{{\rm sim}} \lesssim 10^{14} \rm{g~cm}^{-3}$, cold and warm matter streams alternate.
In the following sections, we explain how these data are used to post-process the original simulations. 

\subsection{EOS modelling \label{sec:eos_modelling}}

In our analysis, we considered baryons, massive leptons, photons, and neutrinos as the relevant \ac{dof} to describe matter and radiation in BNS mergers. Baryons, massive leptons, and photons are assumed to be in thermal and nuclear statistical equilibrium everywhere inside the remnant. Nuclear statistical equilibrium is reached when strong and electromagnetic reactions are in equilibrium, and the nuclear composition is provided by the minimum of the free energy. For astrophysically relevant conditions, this is verified as long as $k_{\rm B}T \gtrsim 0.6~{\rm MeV}$ \citep[e.g.][]{Hix99ApJ}.
In astrophysical plasma, the negative net charge of massive leptons is balanced by the positive charge of protons.
In the presence of more than one massive lepton, charge neutrality implies:
\begin{equation}
Y_p = \sum_{l=e,\mu,\tau} Y_l \ , 
\end{equation}
where $Y_p = n_p/n_b$ is the proton fraction; $n_p$ is the proton number density; $Y_l = (n_{l^-} - n_{l^+}) / n_b$ is the net fraction of the massive lepton $l$; and $e$, $\mu$, and $\tau$ stand for electron, muon, and tauon, respectively.

Every relevant species was characterised by its own EOS. These EOSs can be expressed in terms of any thermodynamical potential, such as the Helmholtz free energy $F$ with the volume $V$, the number of particles $N$, and the temperature $T$, as independent variables. 
Neglecting the modifications of the thermodynamical potentials due to the interaction between different dof, the total free energy was given by
\begin{equation}
    F_{\rm tot} = \sum_{i = b, l, \gamma, \nu} F_i \ ,
\end{equation}
where the subscript $i = \{b, l, \gamma, \nu\}$ indicates baryons, massive leptons, photons, and neutrinos, respectively. 
All the thermodynamic variables, such as the chemical potentials, $\mu_i$, the energy densities, $e_i$, and the pressures, $P_i$, were derived from $F_i$ at fixed temperature $T$, number of particles $N_i$, and volume $V$ according to the standard rules of the canonical ensemble.
Hence, the total energy density and pressure were simply given by
\begin{equation}
    e_{\rm tot} = \sum_{i = b, l, \gamma, \nu} e_i \; , \;
    P_{\rm tot} = \sum_{i = b, l, \gamma, \nu} P_i \ .
\end{equation}
In the following sections, we provide a detailed description of the \acp{EOS} used in this work for each species.\\

\subsubsection{Baryons}\label{sec:baryons} 
Our knowledge of the baryonic matter EOS is still affected by large uncertainties. Therefore, to bracket possible uncertainties in this work, the baryonic contribution was taken into account via three different nuclear EOSs in tabulated form corresponding to the ones used in the simulations presented in \refsec{sec:simulation_sample}: the BLh \citep{bombaci18,logoteta21}, the DD2 \citep{typel10,hempel10}, and the SFHo \citep{steiner13}.\footnote{The corresponding \ac{EOS} tables are publicly available from the CompOSE repository \url{https://compose.obspm.fr/}.} 
We note that we did not consider the possible formation of hyperons \citep{PhysRevC.85.055806,fortin_oertel_providencia_2018} or a phase transition to quark matter \citep{Weissenborn:2011qu, Klahn:2013kga, Chatterjee:2015pua, Bombaci:2016xuj, Logoteta:2019utx, Logoteta:2021iuy}. 
The BLh EOS is a microscopic EOS constructed following the framework of the Brueckner-Bethe-Goldstone many-body theory extended at finite temperatures within the Brueckner-Hartree-Fock approximation. 
In contrast, the modelling of nuclear interactions in DD2 and SFHo is based on a relativistic mean field theory. 
For each nuclear EOS, the independent variables are $(n_b, T, Y_p)$. The range of $(n_b, T, Y_p)$ for each EOS table is reported in \reftab{tab:eos_prop}. The three EOSs are in reasonable agreement with the experimental constraints on the properties of nuclear matter at $n_0$ and $T=0$ (for experimental constraints, see \cite{shlomo06, Danielewicz:2013upa, oertel17, Drischler:2017wtt}). 
Moreover, all three EOSs satisfy to a good extent present astrophysical constraints on the NS maximum mass $M_{\rm max} > 2.01 M_\odot$ and NS radius \citep{antoniadis13, lattimer14, oertel17, Riley:2019yda, Miller:2019cac, raaijmakers21, Miller2021ApJL}.\footnote{We note, however, that the dimensionless tidal deformability predicted by DD2 for GW170817-like events is $ > 800$, which is in possible tension with constraints derived from GW170817 \cite[e.g.][]{abbott17_prl, Abbott:2018wiz, Breschi:2021}.} In \reftab{tab:eos_prop}, we report for each \ac{EOS} the values of the nuclear saturation density, the maximum mass $M_{\rm max}$ of a cold non-rotating NS, the corresponding radius $R_{\rm max}$, the radius $R_{1.4}$ of a $1.4 M_{\odot}$ \ac{NS}, and the dimensionless tidal deformability $\tilde{\Lambda}$ \citep[e.g.][]{Hinderer2008,damour12, favata14} for GW170817 targeted binaries.
We note that SFHo supports a smaller maximum NS mass with a smaller radius. More generally, SFHo produces more compact NSs compared to the other two EOSs. In the case of BLh, it predicts slightly larger $M_{\rm max}$ and $R_{\max}$, but the values are still close to the ones from SFHo. Conversely, DD2 exhibits significantly larger values for $M_{\rm max}$ and $R_{\rm max}$ and a smaller compactness. 
Despite the fact that the maximum NS properties of SFHo and BLh are relatively close, it is worth noting that for other properties at densities equal to several times $n_0$, SFHo and DD2, regarding the symmetry energy, and DD2 and BLh, for incompressibility, are more similar.
With this selection of baryonic EOSs, we could explore very different microphysical descriptions and a wide range of thermodynamical conditions and macroscopic properties of merging BNSs.

\begin{table*}
  \caption[]{Properties of nuclear EOS tables used in this work and of cold, spherically symmetric \acp{NS} for each nuclear EOS. The notations $n_b$, $T$, and $Y_p$ give the range of baryon number density, temperature, and proton fraction, respectively, in each EOS table. The nuclear saturation density is given under $n_0$; $M_{\rm max}$ and $R_{\rm max}$ are the gravitational mass and radius, respectively, of the maximum \ac{NS}; $R_{1.4}$ is the radius of a $1.4 M_{\odot}$ \ac{NS}; and $\tilde{\Lambda}$ is the reduced, dimensionless tidal deformability for GW170817 targeted binaries with mass ratio in the range $1.00 - 1.43$.}
  \label{tab:eos_prop}
  \centering
  \begin{tabular}{ccccccccc}
  \hline \\ [-1.5ex]
  EOS & $n_b$ &$T$ &$Y_p$ &$n_0$ & $M_{\rm max}$ & $R_{\rm max}$ & $R_{1.4}$ & $\tilde{\Lambda}$\\
    & $[{\rm fm}^{-3}]$ & [MeV]& [-]& $[\text{fm}^{-3}]$ & [$M_\odot$] & [km] & [km] & [-]\\
  \hline \\ [-1.5ex]
  BLh  &$10^{-12} - 1.2$ & $0.1 - 158.5$ & $0.01 - 0.6$& $0.171$ & $2.10$ & $10.4$ & $12.4$ & $510 - 513$\\
  DD2 & $10^{-12} - 10$ & $0.1 - 158.5$ & $0.01 - 0.6$ & $0.149$ & $2.42$ & $11.9$ & $13.2$ & $779 - 809$\\
  SFHo & $10^{-12} - 10$ & $0.1 - 158.5$ & $0.01 - 0.6$ & $0.158$ & $2.06$ & $10.3$ & $11.9$ & $392 - 393$\\
  \hline \\ [-1.5ex]
  \end{tabular}
\end{table*}

\subsubsection{Massive leptons and photons} 
We treated massive leptons as ideal Fermi gases in thermal equilibrium. In particular, we incorporated electrons and positrons $e^\pm$ as well as muons and anti-muons $\mu^\pm$, while we neglected tauons and anti-tauons $\tau^{\pm}$ due to their large mass compared to the typical temperatures and Fermi energies in the system.
The EOS of a non-interacting gas\footnote{We note that the electromagnetic interaction of negatively charged leptons with positively charged ions is however taken into account by applying a correction to the lepton chemical potential, as prescribed in \citet{hempel10}.} of fermions at a finite temperature is well known in the literature, and it provides the expressions for the number density $n_{l^\pm}$, the specific energy density $e_{l^\pm}$, and the pressure $P_{l^\pm}$ \citep{bludman77}:
\begin{equation}\label{eq:n_l}
     n_{l^\pm} = K_{l} \theta_{l}^{3/2} \left[ F_{1/2} (\eta'_{l^\pm}, \theta_{l}) + \theta_{l} F_{3/2} (\eta'_{l^\pm}, \theta_{l})\right]
\end{equation}
\begin{equation}
    e_{l^\pm} = K_{l} m_{l} c^2 \theta_{l}^{5/2} \left[F_{3/2}(\eta'_{l^\pm},\theta_{l}) + \theta_{l} F_{5/2} (\eta'_{l^\pm},\theta_{l})\right] + m_l c^2 n_{l^{\pm}}
\end{equation}
\begin{equation}
    P_{l^\pm} = \dfrac{K_{l} m_l c^2}{3} \theta_l^{5/2} \left[ 2 F_{3/2} (\eta'_{l^\pm},\theta_{l}) + \theta_{l} F_{5/2} (\eta'_{l^\pm},\theta_{l}) \right]
,\end{equation}
where $m_l$ is the lepton mass and $K_l$ is a constant
\begin{equation}
    K_l = 8 \sqrt{2} \pi \left(m_l c^2/h c\right)^3 \ ,
\end{equation}
$\theta_l$ and $\eta'_{l^\pm}$ are the relativity and the non-relativistic degeneracy parameters, respectively,
\begin{equation}
\theta_l = \dfrac{k_B T}{m_l c^2}   \quad , \quad \eta'_{l^\pm} = \dfrac{\mu_{l^\pm}-m_l c^2}{k_B T} \ ,
\end{equation}
while $F_k(\eta'_{l^\pm},\theta_l)$ are the generalised Fermi functions of order $k$. At thermal equilibrium, the degeneracy parameter of anti-particles $\eta'_{l^+}$ is related to that of particles $\eta'_{l^-}$ \citep{timmes99}:
\begin{equation}
\eta'_{l^+} = - \left( \eta'_{l^-} + 2 \dfrac{m_l c^2 }{k_B T} \right) \ .
\end{equation}
By inverting \refeq{eq:n_l}, the term $\eta'_{l^-}$ can be expressed as a function of $(n_b, Y_l, T)$, as was suggested in \citet{timmes99}.

Photons form an ideal Bose gas in thermal equilibrium with matter and with zero chemical potential, $\mu_\gamma = 0$.
The resulting EOS of photons depends only on $T$:
\begin{equation}\label{eq:eos_photons}
    n_{\gamma} = \frac{16 \pi k_{\rm B}^3 \zeta(3)}{(hc)^3} T^3 \; , \;
    e_{\gamma} = \frac{8 \pi^5 k_{\rm B}^4}{15~(hc)^3} T^4 \, ,
\end{equation}
and $P_{\gamma} = e_{\gamma}/3$.\\

\subsubsection{Trapped neutrinos}\label{sec:trapped_nu}
Based on the discussion in \refsec{sec:analytical_estimates} and as typical neutrino energies inside the remnant are much larger than any neutrino mass, we modelled the trapped neutrino component as a massless Fermi gas in weak and thermal equilibrium with matter and radiation. Accordingly, the particle number density $n_{\nu}$, the energy density $e_{\nu}$, and the pressure $P_{\nu}$ of any flavour neutrino $\nu$ were given by
\begin{equation}
  n_{\nu} = \dfrac{4 \pi}{(h c)^3} \left(k_B T \right)^3 F_2(\eta_{\nu}) e^{-(\rho_{\rm lim}/\rho)} \, , 
\end{equation}  
\begin{equation}  
  e_{\nu} = \dfrac{4 \pi}{(hc)^3} (k_B T)^4 F_3(\eta_{\nu}) e^{-(\rho_{\rm lim}/\rho)} \, ,
\end{equation}
\begin{equation}
  P_{\nu} = e_\nu/3 \, , 
\end{equation}
where the exponential factor $e^{-(\rho_{\rm lim}/\rho)}$ was introduced in order to model the fading of the trapped component at $\rho \lesssim \rho_{\rm lim}$ (see also \citealt{Kaplan2014ApJ} and \citealt{perego19} for a similar choice). The degeneracy parameter of anti-neutrinos was fixed by $\eta_{\bar{\nu}} = -\eta_{\nu}$ because of thermal equilibrium.
Furthermore, $\eta_{\nu_\tau} = 0$ since the fraction of tauons is negligible, while weak equilibrium defined the degeneracy parameters of electronic and muonic neutrinos:
\begin{eqnarray}\label{eq:degeneracy_parameter}
\eta_{\nu_l} = \eta_p'(n_b, Y_p, T)-\eta_n'(n_b,Y_p,T)+\eta_{l^-}'(n_b,Y_l,T)+\nonumber\\
 +\dfrac{(m_p-m_n+m_l) c^2}{k_B T} \hspace{1cm} (l = e, \mu) \, ,
\end{eqnarray}
where $\eta_n'$ and $\eta_p'$ are the non-relativistic degeneracy parameters of neutrons and protons, respectively. Accordingly, the EOS of electron (muon) neutrinos depends on $T$, $Y_p$, and $Y_e$ ($Y_\mu$), while the EOS of tauon neutrinos depends only on $T$ and does not distinguish between $\nu_\tau$ and $\bar{\nu}_\tau$. The fraction of neutrinos and their mean energy were defined as $Y_\nu = n_\nu / n_b$ and $E_\nu = e_\nu / n_\nu$, respectively.\\

From the previous discussion, if all the species are assumed to be in thermal and reaction equilibrium, the set of variables needed to describe the complete \ac{EOS} is given by $(n_b, T, Y_p, Y_e, Y_\mu$). However, the charge neutrality of \ac{NS} matter constrains the proton fraction $Y_p = Y_e + Y_\mu$, reducing the set of independent variables to $(n_b, T, Y_e, Y_\mu)$. 


\subsection{The post-processing technique}\label{sec:post_proc}

Based on the simulation outcome discussed in \refsec{sec:simulation_sample}, we estimated the amount and the impact of muons in post-processing  as well as their influence on the trapped neutrino properties. In particular, we were interested in the implications for the remnant on a timescale $\Delta t \sim 5 - 15 \rm{ms}$ after the merger and in the density region $\rho \gtrsim 10^{13} \rm{g~cm}^{-3}$ where the approximation of weak and thermal equilibrium better applies. The absence of muons inside the original simulations posed a significant issue and required some modelling (shown in \refsec{sec:pp_modeling}) before the post-processing technique could be applied (discussed in \refsec{sec:pp_algorithm}).

\subsubsection{Modelling}\label{sec:pp_modeling}
\begin{figure*}[h]
\includegraphics[scale=0.67]{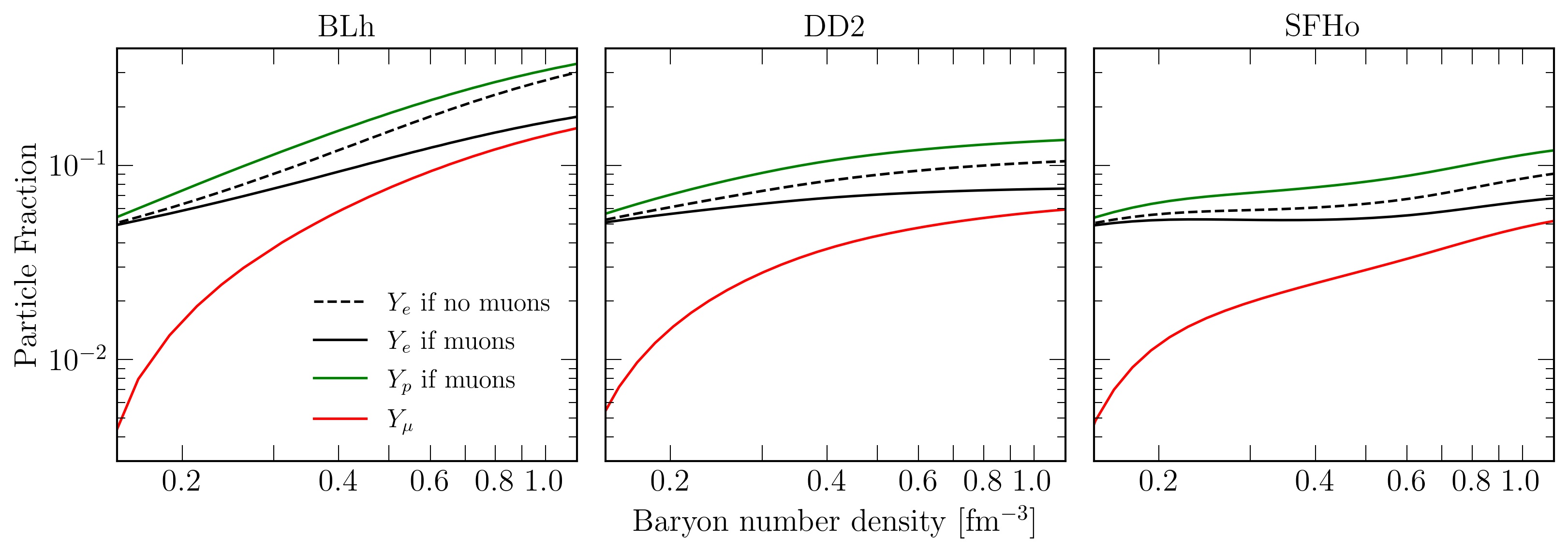}
\caption{\label{fig:2} 
Particle fractions computed at $T = 0$ in neutrinoless $\beta-$equilibrium for three different EOS: BLh (left), DD2 (centre), and SFHo (right). Black lines show the electron fraction computed in two different cases: equilibrium with muons (solid lines) or without muons (dashed lines). The green line corresponds to the fraction of protons at equilibrium with muons, while the red line corresponds to the muon fraction. We note that when muons are not included, the proton fraction coincides with the electron one (dashed black line).}
\end{figure*}

We modelled the \ac{NS} fluid elements before the merger as being composed of neutrons, protons, electrons, and muons in cold neutrinoless weak equilibrium:
\begin{equation}\label{eq:cold_eq}
  \begin{cases}
  \mu_e (T = 0) &= \mu_{n}(T = 0) + \mu_{p} (T = 0) \\
  \mu_\mu (T = 0) &= \mu_e (T = 0) \ .
 \end{cases} 
\end{equation}
In \reffig{fig:2}, we show the fractions of electrons, muons, and protons as a function of $n_b$, obtained by solving \refeq{eq:cold_eq}. At $n_b \approx 0.3 \text{ fm}^{-3} \sim 2 \ n_0$, the muon fraction is approximately 40 \% of the electron fraction for all the considered \acp{EOS}, and it increases monotonically with $n_b$. For comparison, we also show the electron fraction calculated by assuming $\beta$-equilibrium without muons (see dashed black line in \reffig{fig:2}), which corresponds to the initial conditions used in the simulations presented in \refsec{sec:simulation_sample}. A common feature of all three models is that when muons are included, the electron fraction becomes smaller, while the proton fraction tends to increase, that is, $Y_p = Y_e + Y_\mu$. In particular, the system lowers the Fermi energy of electrons by converting electrons into muons up to the point where the chemical potentials of electrons and muons are equal, balancing the difference between the neutron and proton chemical potentials. Moreover, when muons are included, neutrons can turn into protons plus muons via Urca processes \citep{leinson02, Yakovlev:2004iq, Potekhin:2015qsa}. The three \acp{EOS} show pronounced differences in the density behaviour of the electron fraction. In particular, BLh predicts a higher $Y_p$ at large densities compared to DD2 and SFHo. Such differences are mainly due to the different density behaviours of the symmetry energy in the three models. Symmetry energy is indeed the principal source that determines matter composition at a given density and temperature. In the following paragraphs, we refer to the particle fractions of the cold \acp{NS} obtained by solving \refeq{eq:cold_eq} as $\tilde{Y}_e(n_b)$ and $\tilde{Y}_\mu(n_b),$ which are represented as solid lines in \reffig{fig:2}, and to the net electron fraction computed in cold weak equilibrium without muons (see dashed black line in \reffig{fig:2}) as $\tilde{Y}_{e,0}(n_b)$.

During a merger, the density, temperature, and composition of the fluid elements change because of matter compression and decompression, and neutrinos are produced. If a fluid element is in good approximation within the last scattering neutrino surface (i.e. $\rho \gtrsim \rho_{\rm lim}$), a trapped neutrino gas is formed in weak and thermal equilibrium with matter. In these conditions, it is convenient to introduce for each fluid element the electron, muon, and tauon lepton numbers
\begin{equation}
    Y_{l, j} = Y_{j} + Y_{\nu_j} - Y_{\bar{\nu}_j} \qquad j = \{ e, \mu, \tau \}, 
\end{equation}
and the total fluid energy
\begin{equation}
  e_{\rm tot} = \sum_{i} e_i \qquad i = \{b, e^\pm, \mu^\pm, \gamma, \nu_j, \bar{\nu}_j \} \ .
\end{equation}
The central part of the remnant, characterised by $\rho \gtrsim 10^{14} {\rm g~cm}^{-3}$, is formed by the fusion of the two \acp{NS} cores, and it is characterised by fluid elements whose density is significantly in excess of the neutrino surface density during their evolution.
Such fluid elements are not affected by shocks and significant neutrino emission. In particular, as discussed in \refsec{sec:analytical_estimates}, the neutrino diffusion timescale, $t_{{\rm diff}}$, is in the order of seconds; thus $t_{{\rm diff}} \gg \Delta t \sim 5 - 15~\rm{ms}$ (see \refeq{eq:diff_NS}). So these fluid elements conserve their electron and muon lepton fractions over $\Delta t$, which are ultimately equal to the electron and muon lepton fractions set by the initial cold neutrinoless equilibrium, \refeq{eq:cold_eq}.\footnote{At high density, neutrino oscillations are suppressed \citep[see e.g.][]{Richers2019}
so that electron and muon lepton fractions are separately conserved.} A close inspection of the thermodynamic conditions experienced by matter in a \ac{BNS} merger event and of their corresponding evolution \citep{perego19} revealed that the conservation of the lepton numbers is well realised for matter originally inside the outer core of the two cold \acp{NS}, that is, at $\rho \gtrsim 10^{14} \text{g~cm}^{-3}$ also during the inspiral. 
In contrast, most of the matter at a lower density inside the merger remnant, $\rho < 10^{14} {\rm g~cm}^{-3}$, originates from the decompression of matter originally around $\rho \sim 10^{14} {\rm g~cm}^{-3}$ inside the two merging \acp{NS}. In this case, matter is significantly heated by shocks and compression such that the electron and muon lepton numbers of each fluid element become considerably altered. In this density domain, the electron fraction evolved through the numerical relativity simulations is a better proxy of $Y_{l,e}$: $ Y_{l,e} \sim Y_e^{{\rm sim}} $.
The muon lepton fraction is instead approximately given by
$Y_{l,\mu} \sim 0.01 \ , $ which corresponds to the typical muon fraction in the original \acp{NS} at the edge of the outer core, that is, around and just below saturation density (see \reffig{fig:2}).

Concerning the relativistic internal energy, no significant energy leaks out in the form of neutrinos from the fluid elements until $t_{{\rm diff}} \gg \Delta t$ in neutrino trapping conditions. Thus, the evolution of $e_{\rm sim}$ inside the simulation is also expected to reproduce in good approximation the evolution (i.e. the variation) of the total internal energy once muons and neutrinos are present. However, the presence of physical muons inside the two cold \acp{NS} alters the absolute value of the relativistic internal energy due to their rest mass contribution. Nevertheless, this contribution remains approximately constant in time as long as the bulk of the muons present in the remnant comes from the two cold \acp{NS}.

\subsubsection{Algorithm}\label{sec:pp_algorithm}
Based on the above considerations, we designed the following post-processing algorithm.
We considered the time snapshots of the simulations listed in \reftab{tab:sim_sample} in the time interval $\sim 5 - 15$ ms. From each snapshot, we read the full 3D profile of the baryon number density $n_b^{{\rm sim}}$, the net electron fraction $Y_e^{{\rm{sim}}}$, and the internal energy density $e_{{\rm sim}}$ on a Cartesian grid $(x, y, z)$ with $-30~{\rm km} \leq x, y, z \leq +30~{\rm km}$. Given $(n_b^{{\rm sim}}, Y_e^{{\rm sim}}, e_{{\rm sim}})$ on the Cartesian grid and for each snapshot, we performed our analysis according to the following steps.\\

Given $Y_e^{{\rm sim}}$, we computed the baryon density $\bar{n}_b$ such that $Y_e^{{\rm sim}} = \tilde{Y}_{e,0}(\bar{n}_b).$ (See \refsec{sec:pp_modeling} and dashed black line in \reffig{fig:2}.) We note that $\bar{n}_b$ corresponds to the original density of the fluid element inside the \acp{NS} during the inspiral, and it is, in general, different from $n_b^{{\rm sim}}$. 
    
We computed the cold equilibrium electron and muon net fractions corresponding to $\bar{n}_b$ (i.e. $\tilde{Y}_e(\bar{n}_b)$ and $\tilde{Y}_\mu(\bar{n}_b)).$ (See \refsec{sec:pp_modeling} and solid black and red lines in \reffig{fig:2}.)

Given the simulation baryon number density $n_b^{{\rm sim}}$, we computed $\rho_{{\rm sim}} = m_{{\rm amu}}~n_b^{{\rm sim}}$. If $\rho_{{\rm sim}} > 10^{14} {\rm g~cm}^{-3}$, then the lepton numbers were separately conserved (see discussion in \refsec{sec:pp_modeling}). Therefore, we identified the electron and muon lepton fractions with
\begin{equation}\label{eq:assign_yl1}
    Y_{l,e} = \tilde{Y}_e (\bar{n}_b) \ \qquad Y_{l,\mu} = \tilde{Y}_\mu (\bar{n}_b) \ . 
\end{equation}
If instead $\rho_{{\rm sim}} < 10^{14} {\rm g~cm}^{-3}$, we imposed
\begin{equation}\label{eq:assign_yl2}
    Y_{l,e} = Y_e^{{\rm sim}} \ \qquad Y_{l,\mu} = 0.01 \ . 
\end{equation}
In accordance with our assumptions, $Y_{l,\tau}=0$ everywhere and at any time.
We note that the proton fraction (which in the simulation is equal to $Y_e^{\rm sim}$) is now larger than $Y_e^{\rm sim}$ since $\tilde{Y}_p = \tilde{Y}_e + \tilde{Y}_\mu$.

Given the simulation energy $e_{\rm{sim}}$, if $\rho_{{\rm sim}} > 10^{14} {\rm g~cm}^{-3}$, we computed the total internal energy density of each fluid element as
\begin{equation}\label{eq:assign_energy1}
    e_{\rm tot} = e_{\rm sim} + \tilde{Y}_\mu(\bar{n}_{b})~ n_b^{\rm{sim}}  m_\mu c^2 \, ,
\end{equation}
while if $\rho_{{\rm sim}} < 10^{14} {\rm g~cm}^{-3}$, we computed
\begin{equation}\label{eq:assign_energy2}
    e_{\rm tot} = e_{\rm sim} + 0.01~n_b^{\rm{sim}}  m_\mu c^2 \, .
\end{equation}
The terms proportional to $n_b^{\rm sim} m_\mu c^2$ correspond to the rest mass energy density of the muons already present in the two cold \acp{NS}.
    
We imposed $n_b = n_b^{{\rm sim}}$, and we solved the system
\begin{equation}\label{eq:post_proc_sys}
\begin{cases}
    Y_{l,e} &= Y_{e} + Y_{\nu_e} (n_b, T, Y_e, Y_\mu) - Y_{\bar{\nu}_e} (n_b, T, Y_e, Y_\mu) \\
    Y_{l,\mu} &= Y_{\mu}+ Y_{\nu_{\mu}} (n_b, T, Y_e, Y_\mu) - Y_{\bar{\nu}_{\mu}} (n_b, T, Y_e, Y_\mu) \\
    e_{{\rm tot}} &= \sum_{i} e_i (n_b, T, Y_e, Y_\mu) 
,\end{cases} 
\end{equation}
with respect to $(T, Y_e, Y_\mu)$, where $i = \{b, e^\pm, \mu^\pm, \gamma, \nu_j, \bar{\nu}_j \}$ and $j=\{ e,\mu,\tau \}$, to find the equilibrium configuration in the presence of (anti)muons and trapped neutrinos.
The (anti-)neutrino fractions, $Y_\nu$, and energies, $e_{\nu}$, are defined in \refsec{sec:trapped_nu}. We note that the condition on the tauon lepton number, $Y_{l,\tau}=0$, is automatically satisfied in the absence of physical tauons and by the resulting $\eta_{\nu_\tau}=0$ assumption. Nevertheless, the contribution of tauon (anti-)neutrinos was taken into account in the calculation of the total energy and pressure.

We computed all the other thermodynamic variables, such as fractions of trapped neutrinos, chemical potentials, and pressures as functions of $(n_b, T, Y_e, Y_\mu)$, as prescribed in \refsec{sec:eos_modelling}. 

We note that in our procedure, we did not prescribe any time evolution. Rather, we post-processed each available time snapshot of the original simulation in the interval $\sim 5 - 15$ ms after merger.
    
\subsubsection{Limits of validity}\label{sec:pp_limits}
Our post-processing analysis allowed us to compute, a posteriori, the thermodynamics of the merger aftermath taking into account the presence of muons and trapped neutrinos. We note that since there are no muons nor any explicit neutrinos in the original simulations, the definition of $Y_{l,e}$, $Y_{l,\mu}$, and $e_{\rm tot}$ based on the simulation outcome cannot be done in a unique and fully consistent way. The choice described in the previous paragraph has the advantage of including the contribution given by the muons already present in the cold \acp{NS}. However, it relies on the assumption that the merger dynamics and $e_{\rm sim}$, once corrected for the muons rest mass energy, are not significantly altered by them. To bracket the uncertainties due to this choice, in \refapp{sec: alternative choice}, we discuss our results for a different choice of $Y_{l,e}$, $Y_{l,\mu}$, and $e_{\rm tot}$ where the initial muon fraction from the cold \acp{NS} is neglected.

In our modelling of neutrinos, we assumed that above $\rho_{{\rm lim},e}$, there is a full Fermi-Dirac distribution of trapped neutrinos. However, neutrinos with energy smaller than $10 - 20$ MeV were not expected to be in equilibrium with matter in this density regime. Moreover, from \refeq{eq:diff_disc} it follows that the diffusion timescale of neutrinos with energy $E_{\nu} \sim 15$ MeV becomes comparable to $\Delta t \sim 10$ ms around $\rho \sim 10^{12} {\rm g~cm}^{-3}$. For this reason, even though we applied our post-processing procedure in the density domain $\rho > \rho_{\text{lim}, e}$, we restricted our analysis of the results to $\rho > 10^{13} \text{g~cm}^{-3}$, where the assumption of neutrino trapping and weak equilibrium is robust, being $t_{\rm diff} \gg t_{\rm dyn}$, and the post-processing approach is more reliable, being $t_{\rm{diff}} \gg \Delta t$. We also note that our modelling of the trapped neutrino component closely follows the one used by \citet{perego19}. Recent numerical results obtained by including neutrino transport through a grey M1 moment scheme in \ac{BNS} merger simulations \citep{Zappa2023} are very consistent with the outcome of the post-processing analysis of \citet{perego19}.

Finally, we highlight that in our procedure, the population of anti-muons is given by a Fermi-Dirac distribution in thermal equilibrium. Therefore, anti-muons arise as a thermal tail at a high enough temperature.


\section{The fraction of muons in the remnant}\label{sec:muons_remnant}

\begin{figure*}
    \includegraphics[scale=0.6]{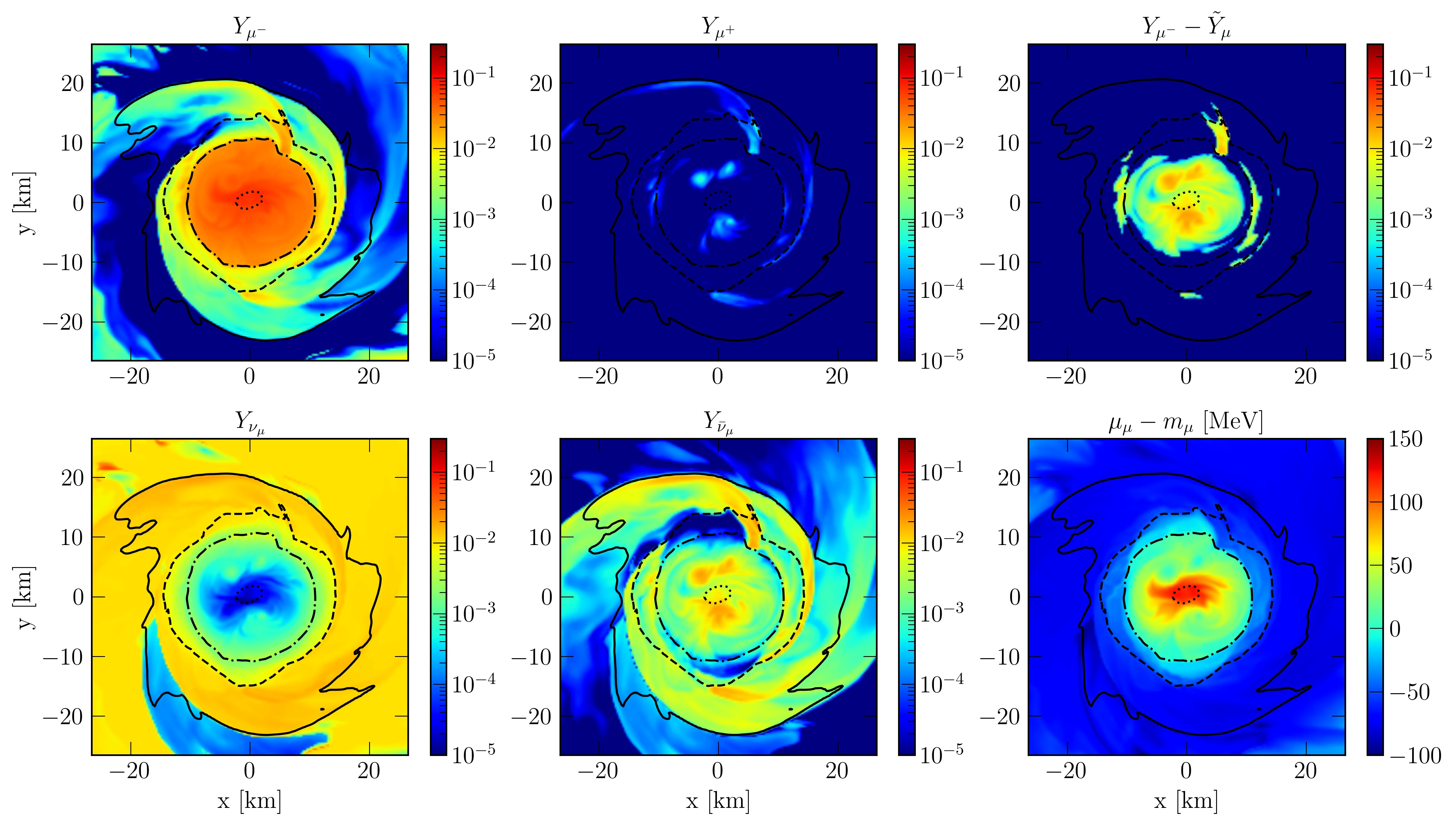}
    \caption{Particle fractions and chemical potential in the muon flavour sector computed by post-processing the simulation (BLh, 1.00) at 4.6 ms after merger. In the upper panels, we show the fraction of $\mu^-$ (left), the fraction of $\mu^+$ (centre), and the difference between the muon fraction and the one inherited from the cold \acp{NS}, that is, $Y_{\mu^-}-\tilde{Y}_{\mu}$ (right). In the lower panels, we show the fraction of $\nu_\mu$ (left), the fraction of $\bar{\nu}_\mu$ (centre), and the muon chemical potential (rest mass subtracted) (right). As in \reffig{fig:1}, the black lines mark isodensity contours.}
    \label{fig:muons_blh}
\end{figure*}

\begin{figure*}
    \centering
    \includegraphics[scale=0.6]{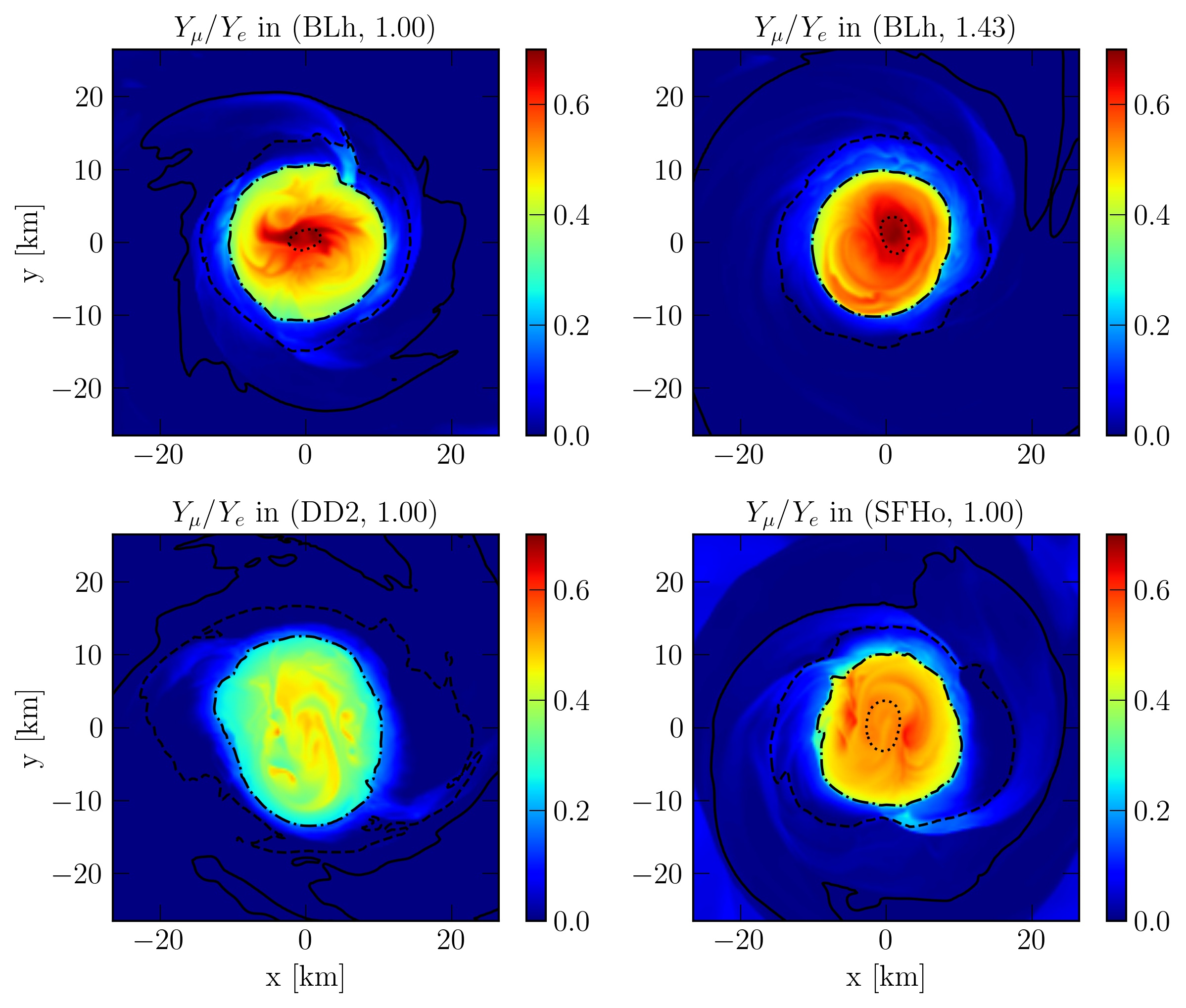}
    \caption{Ratio of the net muon fraction over the net electron fraction, $Y_{\mu}/Y_{e}$, on the equatorial plane obtained by post-processing simulations (BLh, 1.00) (upper left), (BLh, 1.43) (upper right), (DD2, 1.00) (lower left), and (SFHo, 1.00) (lower right) at $5 - 6$ ms after merger. As in \reffig{fig:1}, the black lines mark isodensity contours.}
    \label{fig:ymu_ye_ratio}
\end{figure*}

We started our post-processing analysis from the simulation (BLh, 1.00). 
In \reffig{fig:muons_blh}, we show the fractions of $\mu^{\pm}$, $\nu_\mu$, and $\bar{\nu}_\mu$ as well as the muon chemical potential approximately 5~\rm{ms} after merger. The fraction of muons, $Y_{\mu^-}$, is $\sim 0.01 - 0.07$ in the high-density region characterised by $\rho > 10^{14} \rm{g~cm}^{-3}$, while it decreases down to $\sim 10^{-3} - 10^{-2}$ at the lower density $\rho \sim 10^{13} - 10^{14} \rm{g~cm}^{-3}$. By computing the difference $Y_{\mu^-} - \tilde{Y}_\mu$ (upper-right panel of \reffig{fig:muons_blh} ), we deduced that the bulk of muons comes from the two cold \acp{NS}. This result is consistent with the hypothesis of our post-processing procedure (see \refsec{sec:pp_modeling}). However, a fraction of the muons was created during the merger, corresponding to $\sim 10^{-3} - 10^{-2}$. Muon creation became enhanced in correspondence to the hot spots where $k_{\rm B} T_{\rm sim} \gtrsim 50 \rm{~MeV}$ (see the second panel of \reffig{fig:1}) and at high density, that is, above $10^{15} \rm{g~cm}^{-3}$. By inspecting the change in temperature and particle fractions
\begin{equation}\label{eq:deltaT_deltaY}
    \Delta T = T - T_{\rm sim} \qquad \Delta Y_i = Y_i - Y_i^{\rm sim} ,
\end{equation}
where $T$ and $Y_i$ result from post-processing calculations, while $T_{\rm sim}$ and $Y_i^{\rm sim}$ are read from the simulation, we could guess which were the processes more likely to be involved in muon creation.
In the hot spots, both muons and anti-muons were produced, and the fraction of newly created muons $Y_{\mu^-} \sim 10^{-2}$ exceeded the fraction of anti-muons $Y_{\mu^+} \sim 10^{-3}$ (see \reffig{fig:muons_blh}, upper-centre and right panels).
At the same time, we observed $\Delta T < 0$, $\Delta Y_n < 0$ and $\Delta Y_{e^-} > 0$. We concluded that, on the one hand, thermal processes drive the creation of $\mu^{\pm}$ pairs at the expense of the system's internal energy, and on the other hand, the reduction in the degeneracy of muons due to the high temperature (see \reffig{fig:muons_blh}, lower-right panel) favours the conversion of $n$ into $p + \mu^{-}$, further enhancing the production of $\mu^{-}$. In addition, we found an enhancement in the creation of $e^{\pm}$ pairs and $e^-$, but the amount of newly created $\mu^-$ systematically exceeded that of the newly created $e^-$.
In the core of the remnant, the density of fluid elements was enhanced by matter compression, and $\mu^-$ as well as $e^-$ were created at the expense of the neutrons' degeneracy ($\Delta Y_n < 0$). Finally, at a lower density, $\rho < 10^{14} \text{g~cm}^{-3}$, the initial muon fraction from cold, decompressing \acp{NS} matter was almost entirely converted into muon neutrinos. We note that the values of $Y_{\mu^-}$ are stable in the time domain of our analysis.

Next, we compare the results obtained from the simulation (BLh, 1.00) with the ones from (DD2, 1.00) and (SFHo, 1.00). In the latter cases, muons were also present in the full-density region, $\rho > 10^{13} \rm{g~cm}^{-3}$. However, the muon fraction $Y_{\mu^-} \sim 10^{-3} - 0.05$ presented a slightly smaller maximum with respect to (BLh, 1.00). This is expected since, as discussed in \refsec{sec:pp_modeling}, DD2 and SFHo predict a smaller muon fraction for cold nuclear matter in weak equilibrium (see also \reffig{fig:2}). The amount of muons created during the merger (i.e. $Y_{\mu^-}-\tilde{Y}_\mu)$ is comparable for the three simulations and correlates in the same way with $\Delta T$ and $\Delta Y_i$, defined in \refeq{eq:deltaT_deltaY}. The amount of $\mu^-$ and $\bar{\nu}_\mu$ produced in the centre of the remnant, however, was smaller for (DD2, 1.00) and (SFHo, 1.00) with respect to (BLh, 1.00), with a difference that can reach one order of magnitude if we compare (DD2, 1.00) to (BLh, 1.00). As in the case of the cold, neutrinoless weak equilibrium conditions presented in \reffig{fig:2}, such a difference between the BLh \ac{EOS} and the DD2 \ac{EOS} is mostly due to the larger slope of the symmetry energy in the former case. 
Additionally, the fraction of $\mu^-$ was on average slightly larger in (SFHo, 1.00) than in (DD2, 1.00) because the softer \ac{EOS} SFHo exhibits larger temperatures and densities than the stiffer DD2 \ac{EOS} \citep[see e.g.][]{foucart16a, Sekiguchi:2016, radice18b, perego19}. 

To assess the relevance of muons in the merger remnants, we compared the equilibrium net muon fraction obtained in our post-processing procedure with the equilibrium net electron fraction by analysing the ratio $Y_\mu/Y_e$, as shown in \reffig{fig:ymu_ye_ratio}. We note that for all the simulations, the net muon fraction was at least $30\% - 40\%$ of the net electron fraction. The ratio $Y_\mu/Y_e$ was the largest for (BLh, 1.00), where it can reach a maximum of approximately 0.7, while it reached up to 0.6 and 0.5 for (SFHo, 1.00) and (DD2, 1.00), respectively. We stress, however, that due to the large mass difference between electrons and muons, the amount of both electrons and positrons was largely increased where the temperature was high such that $Y_{e^-} \gtrsim Y_{e^+}$ and $Y_e = Y_{e^-} - Y_{e^+} \sim 0.1$, while $Y_{\mu^-} \gg Y_{\mu^+}$ and $Y_{\mu} \approx Y_{\mu^-}$.

Finally, we post-processed data from the simulation (BLh, 1.43), which had a significantly larger mass asymmetry. We did not observe notable differences while comparing the maximum value of $Y_{\mu}/Y_{e}$ in (BLh, 1.00) and (BLh, 1.43). However, (BLh, 1.43) exhibited a more extended spatial region where $Y_{\mu}/Y_{e} \sim 0.5 - 0.6$ (see \reffig{fig:ymu_ye_ratio}). This region is characterised by a large temperature, $\gtrsim 40$ MeV, because it embeds the core of the secondary \ac{NS}, which was broadened and strongly heated by compression and shocks during the merger.

\begin{figure*}
    \includegraphics[scale=0.55]{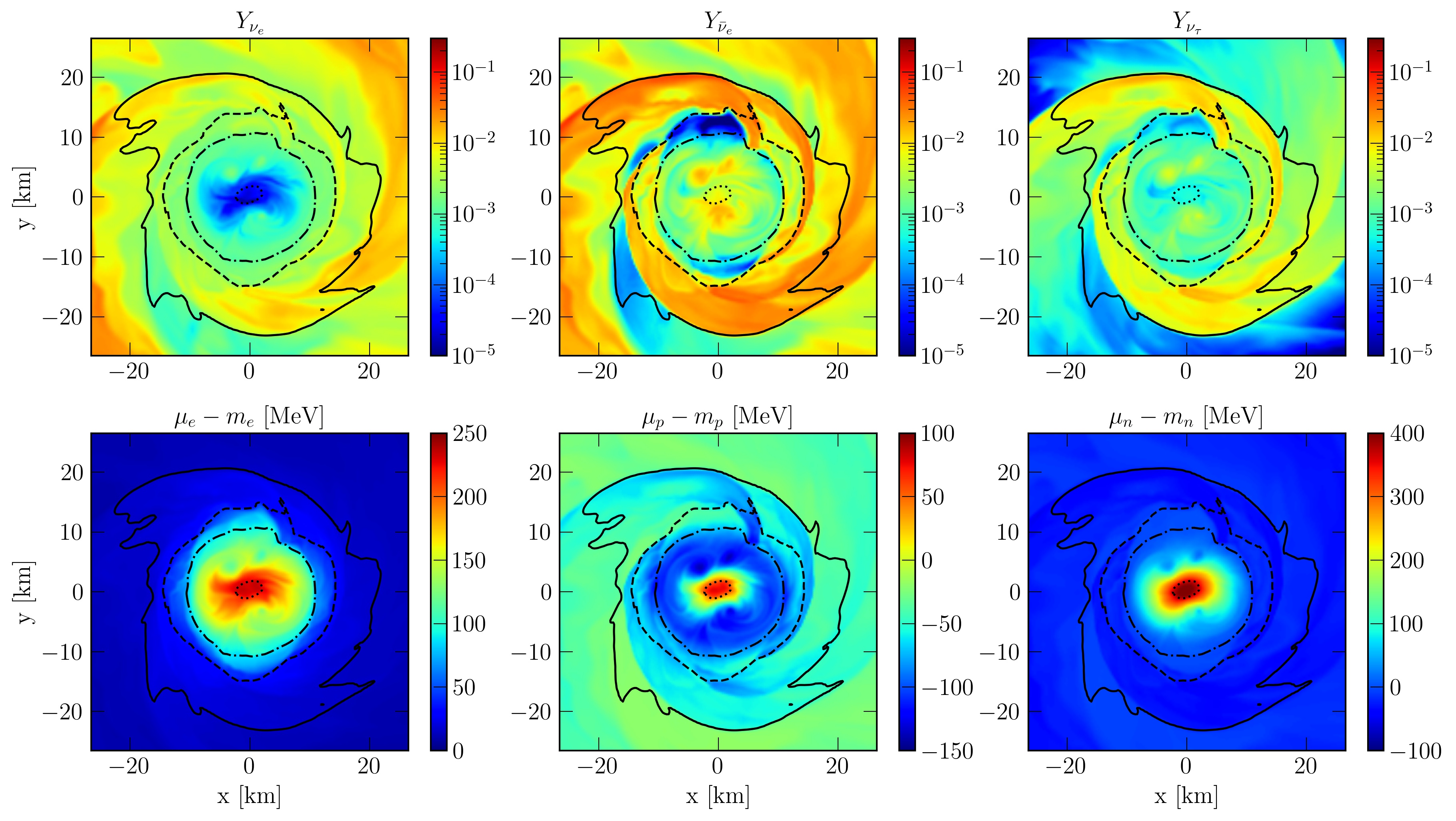}
    \caption{Fractions of $\nu_e$, $\bar{\nu}_e$, $\nu_\tau$, and chemical potentials of electrons, protons, and neutrons computed in post-processing for (BLh, 1.00) at 4.6 ms after merger. In the upper panels, we show the fraction of $\nu_e$ (left), the fraction of $\bar{\nu}_e$ (centre), and the fraction of $\nu_\tau$ (right). In the lower panels, we show the electron chemical potential (rest mass subtracted) $\mu_e - m_e$ (left), the proton chemical potential (rest mass subtracted) $\mu_p - m_p$ (centre), and the neutron chemical potential (rest mass subtracted) $\mu_n - m_n$ (right). We note that the colour bar scales are different for the various chemical potentials. As in \reffig{fig:1}, the black lines mark the isodensity contours.}
    \label{fig:nu_ele_blh}
\end{figure*}

\section{Trapped neutrino properties}\label{sec:nu_trapping}

\begin{figure*}
    \centering
    \includegraphics[scale=0.64]{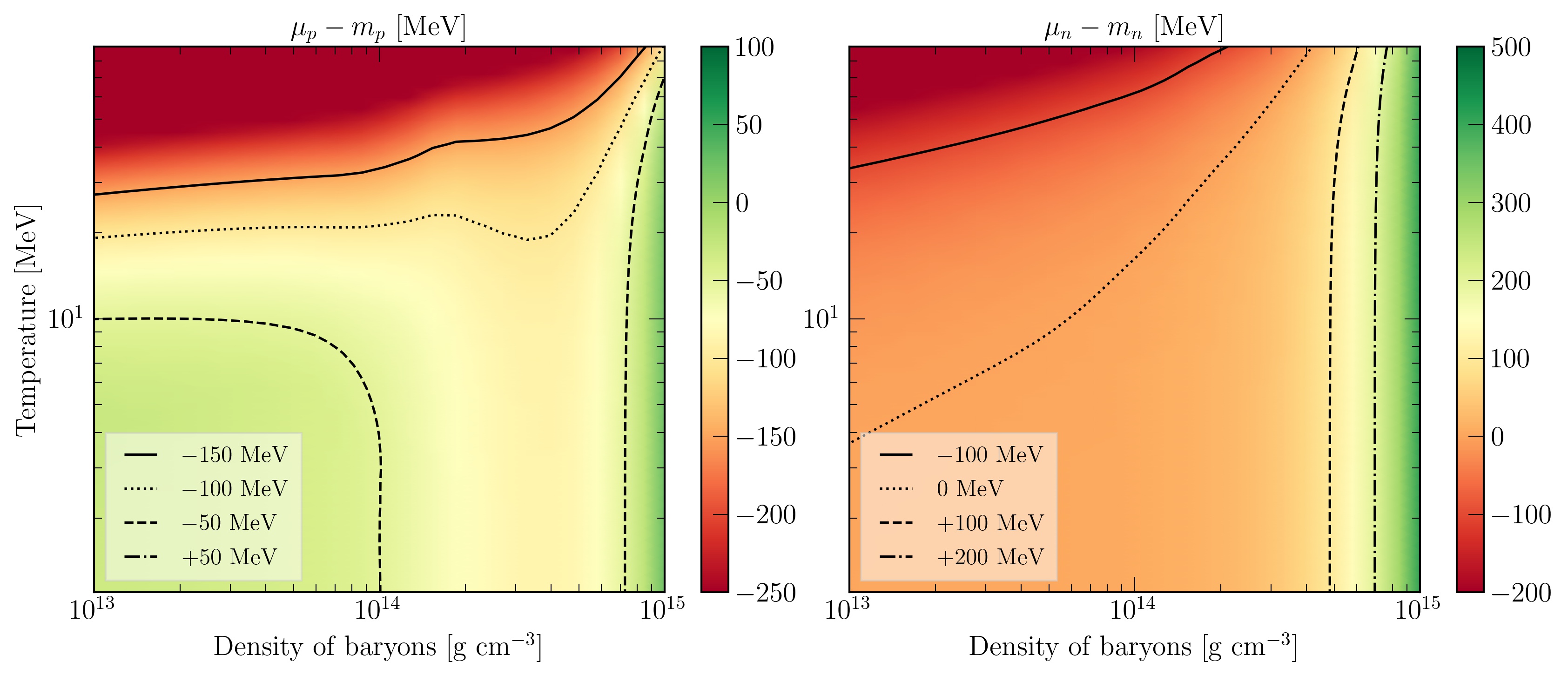}
    \caption{Proton and neutron chemical potentials (rest mass subtracted) of BLh in the density-temperature plane at fixed $Y_p = 0.08$. The proton fraction $Y_p = 0.08$ is the equilibrium proton fraction at 4.6 ms after merger around the point x = 0.64 km, y = 12.76 km, z = 0 km, where $\rho \sim 5 \times 10^{13} \rm{g~cm}^{-3}$, $T_{\rm{sim}} \sim 10$ MeV, and neutrinos dominate, while anti-neutrinos are suppressed (see also \reffig{fig:1}, \reffig{fig:muons_blh} and \reffig{fig:nu_ele_blh}). We notice that the colour bars are scaled differently.}
    \label{fig:mu_baryons}
\end{figure*}

In \reffig{fig:muons_blh} and \reffig{fig:nu_ele_blh}, we show all flavour (anti-)neutrino abundances as well as the chemical potentials of baryons and massive leptons computed by post-processing the simulation (BLh, 1.00) at approximately 5 ms after merger. Trapped gases of (anti-)neutrinos were present in the full-density regime of our analysis, $\rho > 10^{13} \rm{g~cm}^{-3}$, but the abundances and the species hierarchy varied according to the system density and temperature. 

In the very-high-density regime, $\rho \gtrsim 7 \times 10^{14} \rm{g~cm}^{-3}$, $\nu_e$ and $\nu_\mu$ were suppressed ($Y_{\nu_{e, \mu}} < 10^{-4}$), while $\bar{\nu}_e$ and $\bar{\nu}_\mu$ had typical fractions $Y_{\bar{\nu}_{e,\mu}} \sim 10^{-3} - 10^{-2}$, with $\bar{\nu}_\mu$ being slightly more abundant than $\bar{\nu}_e$. We note that the excess of $\bar{\nu}_e$ and $\bar{\nu}_\mu$ exactly compensates for the amount of newly created electrons and muons (see the discussion in the previous section). Among the remaining neutrino species, $\nu_\tau$ and $\bar{\nu}_{\tau}$ were the most abundant, with a fraction $\sim 10^{-4} - 10^{-3}$.

When $\rho \sim 2 - 6 \times 10^{14} \rm{g~cm}^{-3}$, the fractions of neutrinos increased up to $\sim 10^{-3}$, but electron and muon anti-neutrinos still remained the most abundant neutrino species, with $Y_{\bar{\nu}_{e, \mu}} \sim 10^{-3} - 0.02$. All the (anti-)neutrino abundances peaked in the hot spots. However, muon production dominated in this density and temperature regime (see the discussion in the previous section) so that the $\bar{\nu}_\mu$ species was the most abundant, followed by $\bar{\nu}_e$ and then $\nu_\tau$, $\nu_e$, and $\nu_\mu$, in that order. 

At the lower density, $\rho \sim 10^{13} - 10^{14} \rm{g~cm}^{-3}$, all neutrinos and anti-neutrinos were present in warm matter streams characterised by $k_{\rm B} T_{\rm sim} \sim 25$ MeV, with order of magnitude fractions $\sim 10^{-2}$, and the $\bar{\nu}_e$ species dominated, followed in order by $\bar{\nu}_\mu$, $\nu_\tau$, $\nu_\mu$, and $\nu_e$. On the contrary, in cold matter streams with $k_{\rm B} T_{\rm sim} \sim 10 - 15$ MeV, anti-neutrinos were suppressed, and the $\nu_\mu$ species dominated, with a maximum fraction of $\sim 0.01$, followed by $\nu_e$, with $Y_{\nu_e} \gtrsim 10^{-3}$, and $\nu_\tau$, with $Y_{\nu_\tau} \sim 10^{-4}$.

The different abundances and the hierarchy reflect the behaviour of the neutrino chemical potentials at equilibrium (see \refeq{eq:degeneracy_parameter}), which ultimately depend on the \ac{EOS} both directly, through the nuclear interaction, and indirectly, through the different thermodynamic conditions realised in the remnant.
The regions where anti-neutrinos dominated over neutrinos were characterised by a higher density and a higher value of the neutron chemical potential $\mu_n$, which exceeded $\mu_p + \mu_{l^-}$ with $l = \{ e, \mu \}$, as we deduced from \reffig{fig:muons_blh} and \reffig{fig:nu_ele_blh}. In such regions, the density was increased by matter compression, and the fraction of protons increased as well since the lepton number was fixed. This coincided with the enhancement of electrons and muon production at the expense of free neutrons (see discussion in \refsec{sec:muons_remnant}), and the equilibrium condition \refeq{eq:degeneracy_parameter} enforced the creation of the corresponding anti-neutrinos. In contrast, neutrinos dominated in relatively cold matter streams in the outer layers, where the proton chemical potential increased (see \reffig{fig:nu_ele_blh}) so that $\mu_p + \mu_{l^-} > \mu_n$. The enhancement of $\mu_p$ is indeed correlated with the decrease of both density and temperature experienced by the expanding matter in such regions. This is illustrated in \reffig{fig:mu_baryons}, where we plotted $\mu_p$ and $\mu_n$ in the $\rho - T$ plane for $Y_p$ fixed to a relevant value for the analysis (i.e. $Y_p = 0.08$). We note that $\mu_p$ started increasing when $\rho \lesssim 10^{14} \text{g~cm}^{-3}$ and $k_{\rm B} T \lesssim 15$~MeV, while $\mu_n$ decreased in the same $\rho - T$ region. This is because when $\rho \lesssim 10^{14} \text{g~cm}^{-3}$ and $k_{\rm B} T \lesssim 15$~MeV, nucleons start clustering in nuclei so that the chemical potential of free protons is enhanced and the amount of free protons drops \citep[see e.g. Figure 8 of][]{hempel10}. At a fixed lepton fraction and at equilibrium, this results in a conversion of electrons and muons into the corresponding neutrinos. 

\begin{figure*}
    \centering
    \includegraphics[scale=0.6]{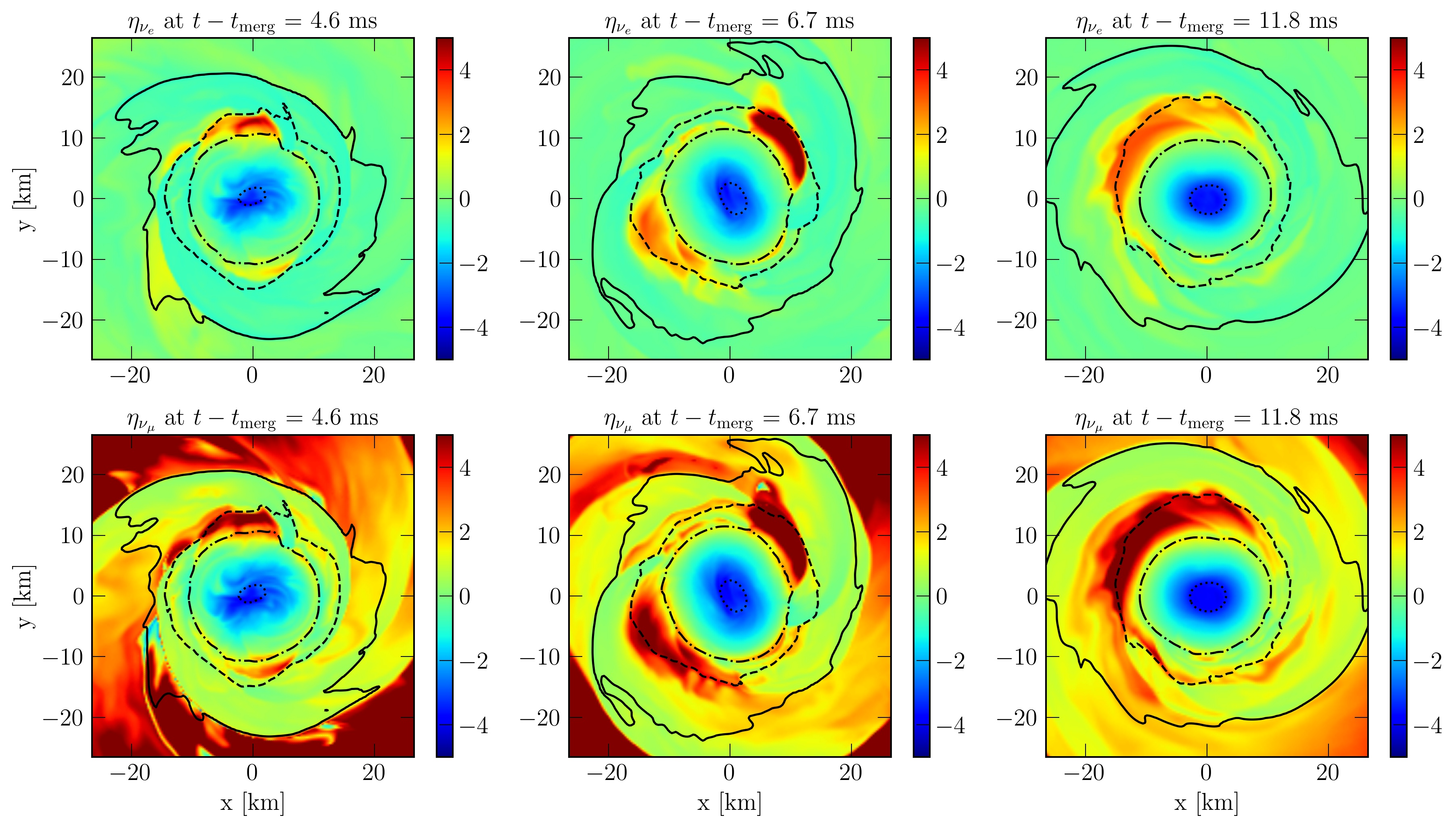}
    \caption{Degeneracy parameter of $\nu_e$ (first row) and $\nu_\mu$ (lower row) on the equatorial plane obtained by post-processing the simulation (BLh, 1.00). Three different time slices are shown corresponding from left to right to 4.6 ms, 6.7 ms, and 11.8 ms after merger. As in \reffig{fig:1}, the black lines mark the isodensity contours as in \reffig{fig:1}.}
    \label{fig:eta_BLh}
\end{figure*}

The properties of the neutrino gases are better characterised in terms of their degeneracy parameters.
In \reffig{fig:eta_BLh}, we show the degeneracy parameter of electron and muon neutrinos computed for three snapshots of (BLh, 1.00): $\sim 5$~ms, $\sim 7$~ms, and $\sim 12$~ms after merger. The spatial distribution of $\eta_{\nu_e}$ and $\eta_{\nu_\mu}$ clearly indicates the presence of trapped degenerate gases of the $\bar{\nu}_e$ and $\bar{\nu}_\mu$ species in the remnant core and of the $\nu_e$ and $\nu_\mu$ species in the outer layers. In this latter region, the trapped gas of electron neutrinos is characterised by a degeneracy parameter $\eta_{\nu_e}$ ranging between three and six, while in the remnant core degenerate electron antineutrinos have $\eta_{\bar{\nu}_e} = - \eta_{\nu_e} \sim 4$. The degenerate gases of muonic neutrinos (in the outer layers) and muonic antineutrinos (in the core) exhibited the same qualitative behaviour as the $\nu_e$ and $\bar{\nu}_e$ species, with a degeneracy parameter $\eta_{\nu_\mu}$ ranging between -4 and 9 so that these muon (anti-)neutrino gases were slightly more degenerate than the electron ones.

The spatial distribution of the abundances as well as the neutrino hierarchy in (DD2, 1.00) and (SFHo, 1.00) are fully analogous to the ones of (BLh, 1.00). Accordingly, we can analyse the differences among the three \acp{EOS} by inspecting the degeneracy parameters. Similarly to (BLh, 1.00), $\bar{\nu}_e$ and $\bar{\nu}_\mu$ formed degenerate trapped gases in the remnant core, while $\nu_e$ and $\nu_\mu$ formed degenerate trapped gases in the cold matter streams of the outer layers. The degeneracy parameter of electron neutrinos, $\eta_{\nu_e}$, ranged between -2 and 6 for both (DD2, 1.00) and (SFHo, 1.00), while in the case of muonic neutrinos, we found $\eta_{\nu_\mu}$ ranging between -2 and 11 for (DD2, 1.00) and between -2 and 12 for (SFHo, 1.00).
In general, the gases of $\bar{\nu}_e$ and $\bar{\nu}_\mu$ are more degenerate for (BLh, 1.00) than for (DD2, 1.00) and (SFHo, 1.00).
The main cause of this resides in the differences among the neutron and the proton chemical potentials, which depend on the selected \ac{EOS} and on the values of $\rho$, $T$, and $Y_p$ in the remnant core. In particular, the average values of $\mu'_n-\mu'_p$ (rest masses subtracted) at a radius of approximately 3 km from the centre of the remnants correspond to $\sim 200$ MeV for DD2, $\sim 245$ MeV for SFHo, and $\sim 325$ MeV for BLh. Hence, the degeneracy of the trapped anti-neutrinos is comparable for DD2 and SFHo but quite different for BLh (see \refeq{eq:degeneracy_parameter}). In this sense, anti-neutrino trapping in the core of the remnant shows a significant and non-trivial dependence on the \ac{EOS} through the nucleons' chemical potentials. If the merger dynamics are significantly affected, this characteristic could become a probe for investigating matter properties in the high-density regime.

Finally, by post-processing the results of the simulation (BLh, 1.43), we observed that the degeneracy of the trapped component of anti-neutrinos was the same as in (BLh, 1.00). However, the trapped neutrino gases at $\rho \sim 10^{13} - 10^{14} \rm{g~cm}^{-3}$ were less degenerate for $q=1.43$, displaying $\eta_{\nu_e} \sim 4$ and $\eta_{\nu_\mu} \sim 7$. This feature depends mostly on the temperature in the density region $10^{13} - 10^{14} \rm{g~cm}^{-3}$, which is systematically larger in the case of $q = 1.43$ compared to $q = 1.00$.

\begin{figure*}
    \centering
    \includegraphics[scale=0.64]{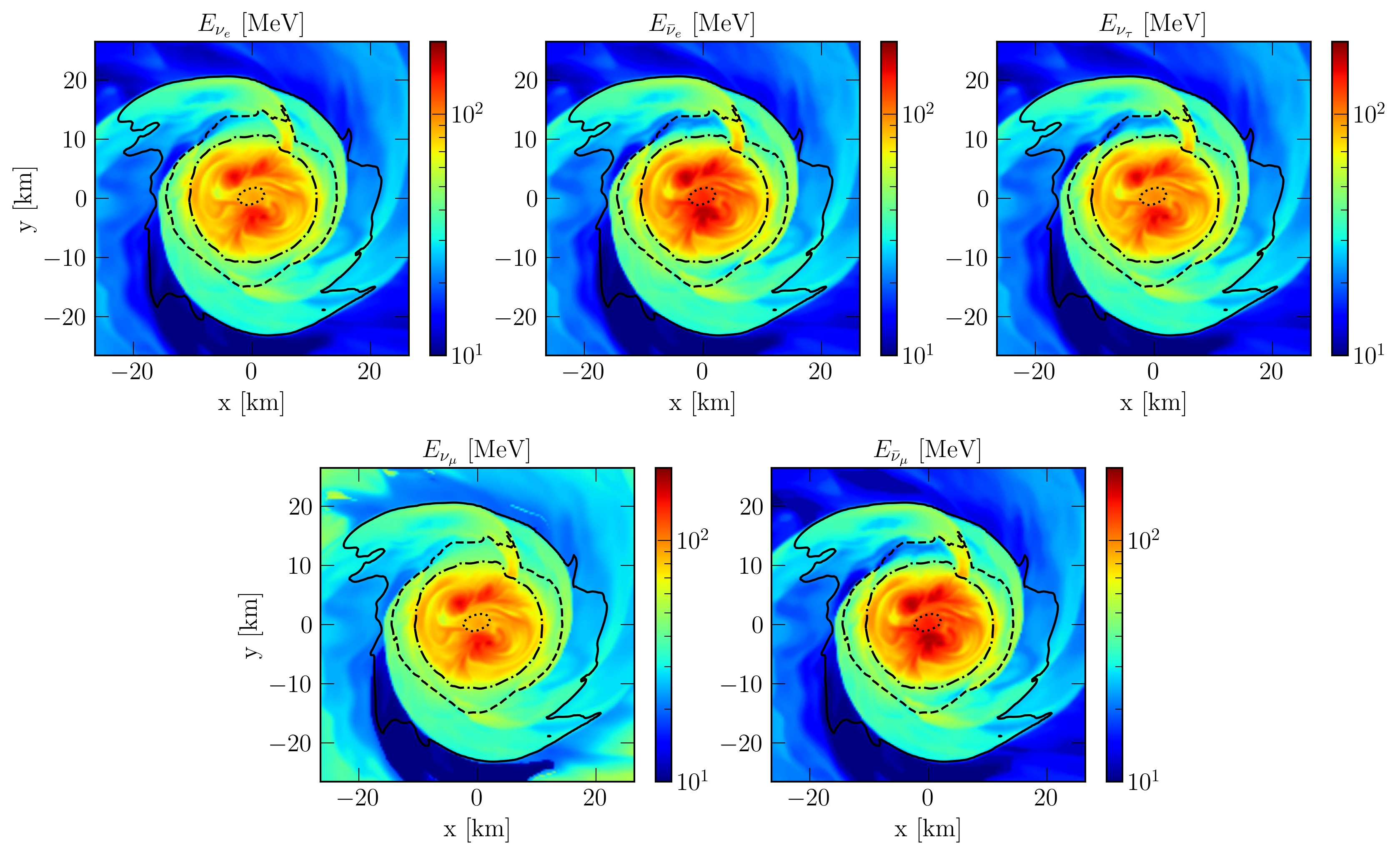}
    \caption{Mean energy $E_{\nu}$ of trapped neutrinos on the equatorial plane computed by post-processing simulation (BLh, 1.00) at 4.6 ms after merger. As in \reffig{fig:1}., the black lines mark isodensity contours.}
 \label{fig:energy_blh}
\end{figure*}

To test the validity of the approximations used in \refsec{sec:analytical_estimates} and to further characterise the properties of neutrinos trapped inside the remnant in the presence of muons for the first time, we computed the neutrino mean energies, $E_{\nu}$, in post-processing. In \reffig{fig:energy_blh}, we show the spatial distribution of $E_\nu$ for the remnant of simulation (BLh, 1.00) at approximately 5 ms after merger. In the density region $\rho > 10^{13} \rm{g~cm}^{-3}$, we found $E_{\nu} > 30$ MeV for all flavour neutrinos, unless a certain flavour was suppressed in a portion of space. Electron and muon anti-neutrinos were the most energetic, with $E_{\bar{\nu}_{e,\mu}} \lesssim 180$ MeV. Then, for $\nu_\tau$ we found $E_{\nu_\tau} \lesssim 170$ MeV, while $\nu_e$ and $\nu_\mu$ were the least energetic, with $E_{\nu_{e,\mu}} \lesssim 160$ MeV. In the case of (DD2, 1.00), the values of $E_\nu$ closely followed the ones discussed for (BLh, 1.00). However, (SFHo, 1.00) exhibited larger energies for all flavour neutrinos, with $E_{\nu_{e,\mu,\tau}} \lesssim 180$ MeV and $E_{\bar{\nu}_{e, \mu}} \lesssim 200$ MeV, due to larger remnant temperatures.

\section{\label{sec:pressure}Changes in the remnant pressure}

\begin{figure*}
    \centering
    \includegraphics[scale=0.55]{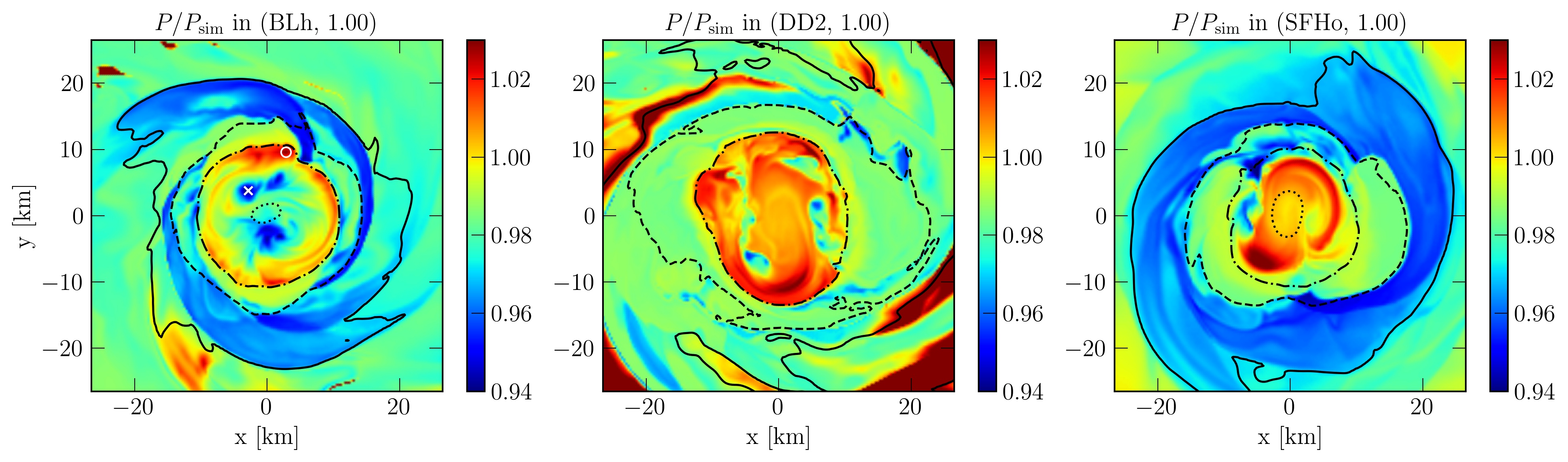}
    \caption{Pressure ratio $P/P_{\rm sim}$ for (BLh, 1.00) (left), (DD2, 1.00) (centre), and (SFHo, 1.00) (right) where $P$ is obtained in post-processing, while $P_{\rm sim}$ is the simulation pressure. For (BLh, 1.00), we explicitly mark the maximum $P/P_{\rm sim} \sim 1.03$ (white circle) and the minimum $P/P_{\rm sim} \sim 0.94$ (white cross) at $\rho > 10^{14} \text{g~cm}^{-3}$. As in \reffig{fig:1}., the black lines mark isodensity contours.  }
    \label{fig:dp}
\end{figure*}

In this section, we investigate how the result of the post-processing, which consistently included muons and trapped neutrinos, modifies the pressure of the remnant with respect to the values extracted from the original simulation.
 In \reffig{fig:dp}, we show the ratio between the pressure computed in post-processing and the simulation pressure, $P/P_{\rm sim},$ for (BLh, 1.00), (DD2, 1.00), and (SFHo, 1.00). Similar to \refsec{sec:muons_remnant} and \refsec{sec:nu_trapping}, we start our discussion with the (BLh, 1.00) simulation. In the time domain of our analysis, the equilibrium pressure computed in post-processing decreased by $6 - 7 \%$ in the region characterised by $\rho \sim 5 \cdot 10^{14} \text{g~cm}^{-3}$, while it increased by a maximum of $3 \%$ around $10$ km from the centre at $\rho \sim 10^{14} \text{g~cm}^{-3}$ compared to the pressure from the original simulation.\\
 
\begin{figure*}
    \centering
    \includegraphics[scale=0.52]{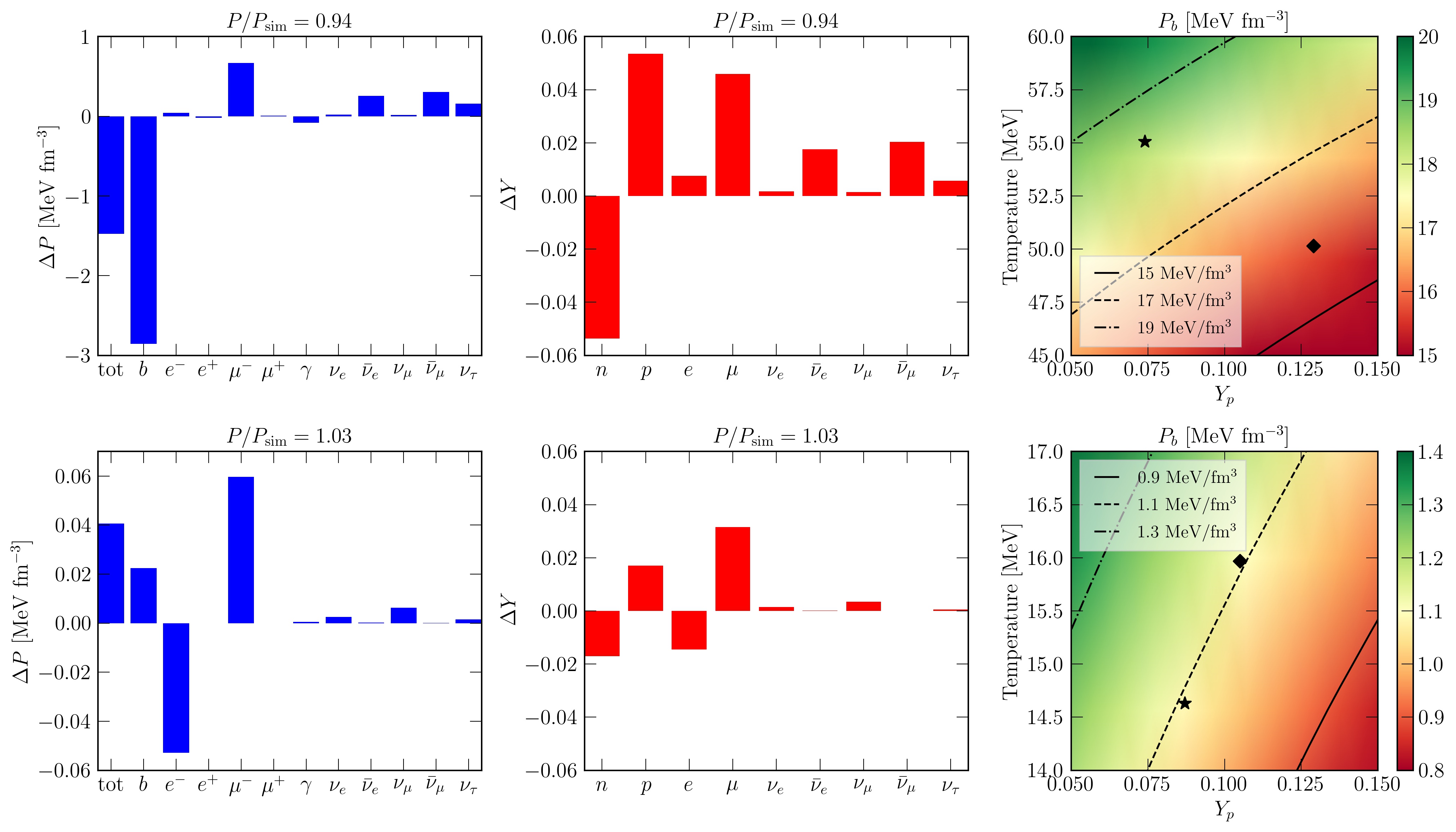}
    \caption{Change of particle pressure and fractions obtained in correspondence to the space-time coordinates explicitly marked in the left panel of \reffig{fig:dp} and corresponding to $P/P_{\rm sim} = 0.94$ (upper row) and $P/P_{\rm sim} = 1.03$ (lower row). The bar plots in blue (left) give the difference between the equilibrium pressure computed in post-processing and in the simulation, $\Delta P_i = P_i - P_i^{\rm ~sim}$, where $i = \rm{tot}, b, e^-, e^+, \mu^-, \mu^+, \gamma, \nu_e, \bar{\nu}_e, \nu_\mu, \bar{\nu}_\mu, \nu_\tau$, and $P_{\rm tot}$ refers to the total pressure. The bar plots in red (centre) show the difference between the particle fractions computed in post-processing and the ones from the simulation, $\Delta Y_i = Y_i - Y_i^{\rm sim}$, where $i = n, p, e, \mu, \nu_e, \bar{\nu}_e, \nu_\mu, \bar{\nu}_\mu$, and $e$ ($\mu$) refers to the net electron (muon) fractions. The colour-coded plots (right) show $P_b$ of BLh as a function of temperature and proton fraction for the same points. The density $\rho$ is fixed to the value obtained in the simulation, that is, $\rho = 4.3 \cdot 10^{14} \rm{g~cm}^{-3}$ (upper row) and $\rho = 1.55 \cdot 10^{14} \rm{g~cm}^{-3}$ (lower row). The black star and the diamond mark are the values from the simulation (before post-processing) and at equilibrium (after post-processing), respectively.}
 \label{fig:inspect_dp}
\end{figure*}

 To understand which processes are responsible for the changes in pressure, we analysed the role of each particle species in detail. In \reffig{fig:inspect_dp}, we have plotted the difference in pressures and particle fractions after muons and neutrinos were introduced
 \begin{equation}
     \Delta P_i = P_i - P_i^{{\rm sim}} \qquad \Delta Y_i = Y_i - Y_i^{{\rm sim}},
 \end{equation}
 for all the particle species at two representative points inside the simulations, explicitly marked in \reffig{fig:dp} and corresponding to $P/P_{\rm sim} \sim 0.94$ and $P/P_{\rm sim} \sim 1.03$, respectively. The drop in pressure was localised inside a hot spot at $k_{\rm B} T_{\rm sim} \sim 55$ MeV, and it was mainly due to a decrease of baryonic pressure. Once muons were included in the microphysics, the system reached a new equilibrium characterised by a smaller neutron fraction; larger proton, electron, and muon fractions; and a smaller temperature (see also discussion in \refsec{sec:muons_remnant}). The decrease in both the neutron fraction and of the temperature lowered the baryonic pressure, and the pressure provided by muons and trapped anti-neutrinos was not large enough to compensate. Accordingly, the total pressure also decreased. This effect is strongly correlated with the enhancement in the production of massive leptons and (anti-)neutrinos at high temperatures.
An increase of pressure occurred at a lower initial temperature, $k_{\rm B} T_{\rm sim} \sim 15$ MeV, and it was correlated with the inclusion of muons from the cold \acp{NS}. As shown in \reffig{fig:2}, including muons in cold neutrinoless $\beta$-equilibrium implies lowering the initial electron and neutron fraction in favour of the muon and proton fractions.
Therefore, the pressure of the system increased because some relativistic particles (electrons) were replaced with non-relativistic particles (muons). An additional contribution came from the increase of the baryon pressure $\Delta P_b = P_b - P_b^{\rm sim} > 0$, where $P_b$ is computed in post-processing while $P_b^{\rm sim}$ is read from the simulation outcome. In general, we would expect $\Delta P_b < 0$ since the neutron fraction was reduced. However, we observed a small increase of temperature $\Delta T = 1.3$ MeV, which was large enough to result in $\Delta P_b > 0$, as shown in the colour-coded plot in the second row of \reffig{fig:inspect_dp}. We note that the increase in temperature was expected because it is the only way the system could simultaneously reduce the number of neutrons and electrons, which are highly degenerate at $k_{\rm B} T_{\rm sim} \sim 15$ MeV and at a fixed internal energy. Interestingly, adding new dof without supplementary repulsive forces did not simply result in a softening of the EOS, as we would have expected, because of the role played by the equilibrium temperature. 

The causes of the changes in pressure in (DD2, 1.00) and (SFHo, 1.00) are the same as in (BLh, 1.00). From our analysis up to 15 ms after merger, we found that $P/P_{\rm sim}$ ranges from 0.94 to 1.05 for (DD2, 1.00) and from 0.93 to 1.04 for (SFHo, 1.00). As shown in \reffig{fig:dp}, (DD2, 1.00) was subjected mostly to a pressure increase, while (SFHo, 1.00) exhibited a pressure increase in the region closer to the core and a pressure decrease in the outer shells. The amount of pressure decrease is comparable to the one found in \citet{perego19}, while the pressure increase is a new prediction. 

Pressure modifications correlate in a complex way with other thermodynamical variables, such as density, electron and muon fractions, temperature, and chemical potentials. However, a more important role is played by temperature. A close inspection of the distribution of $P/P_{\rm sim}$ in the $(\rho_{\rm{sim}}, T_{\rm sim}, Y_e^{\rm sim})$ space revealed that for all the \acp{EOS}, the pressure decrease was more frequent at larger $T_{\rm sim}$. This is because electrons and muons are less degenerate, so the of conversion neutrons into protons is favoured and the baryonic pressure reduces to a greater extent. Regarding the pressure increase, it was frequent at smaller $k_{\rm B} T_{\rm sim} < 20$ MeV because of the dominant role played by the muons coming from the \acp{NS} in colder temperature regimes. We concluded that the remnant formed in (DD2, 1.00) exhibits mostly a pressure increase because its temperature is, on average, smaller compared to (BLh, 1.00) and (SFHo, 1.00).

\begin{figure}
    \centering
    \includegraphics[scale=0.8]{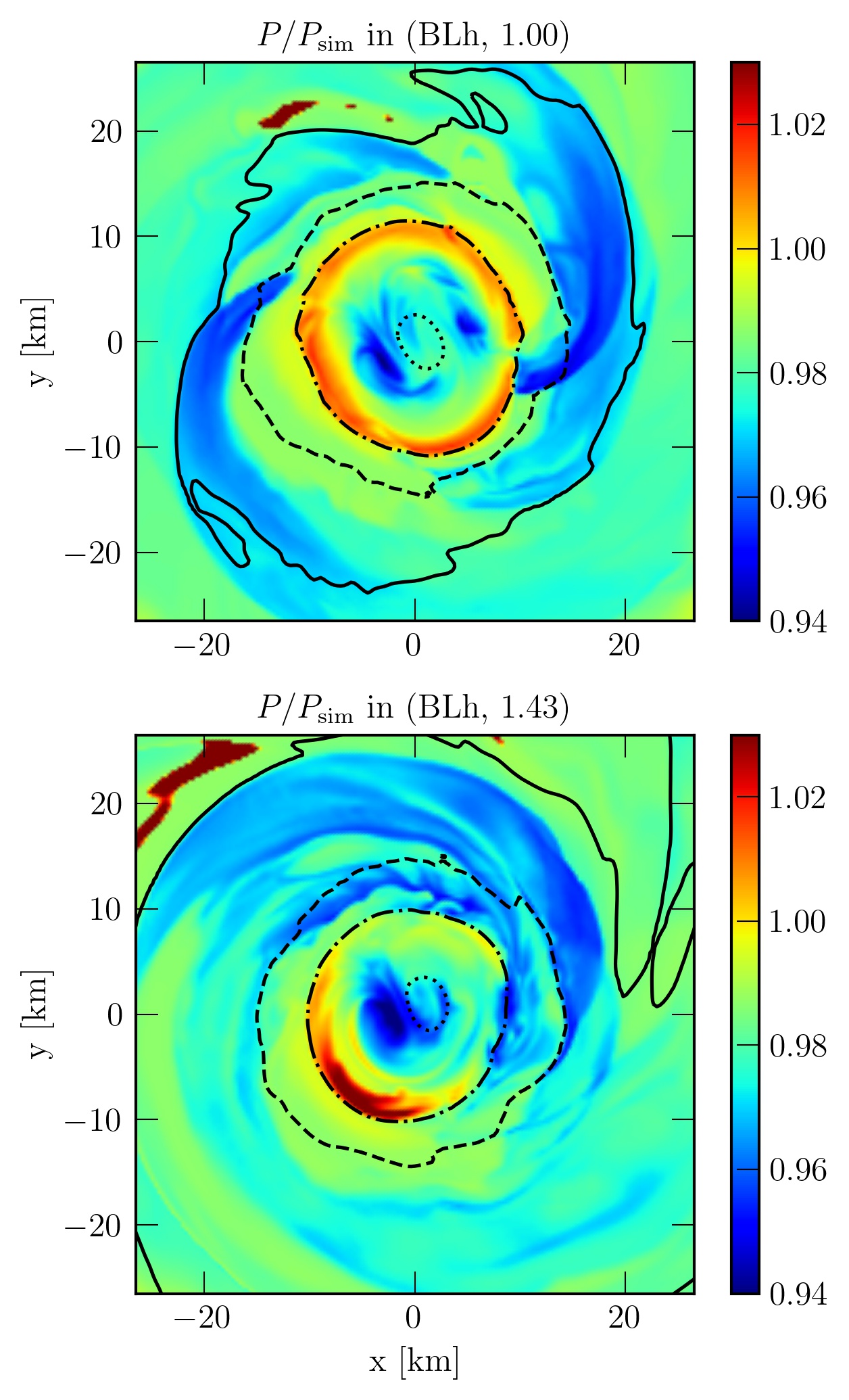}
    \caption{Pressure ratio $P/P_{\rm sim}$ computed by post-processing (BLh, 1.00) (top) and (BLh, 1.43) (bottom) at 6.7 ms after merger. As in \reffig{fig:1}, the black lines mark the isodensity contours.} 
    \label{fig:dp_q}
\end{figure}

We conclude this section by analysing how the pressure is modified in the case of an asymmetric binary mass ratio. If $q = 1.43$, then the ratio $P/P_{\rm sim}$ ranges in the interval $0.93 - 1.05$, showing a pressure decrease compatible with the case where $q = 1.00$ but has a pressure increase that is significantly larger. This behaviour depends mostly on the difference between the electron fractions in the two simulations. In the case of $q = 1.43$, the regions with the highest $P/P_{\rm sim}$ also have a larger $Y_e^{\rm sim}$ compared to $q = 1.00$ and as a consequence a larger $\tilde{Y}_\mu$ (see \reffig{fig:2}). Since the pressure increase is mainly driven by the muon fraction coming from the cold \acp{NS}, a larger $\tilde{Y}_\mu$ results in a larger $P/P_{\rm sim}$. In \reffig{fig:dp_q}, we show the comparison between the pressure ratios for $q = 1.00$  and $q = 1.43$  at approximately 7 ms after merger. We found that the spatial distribution of $P/P_{\rm sim}$ for $q = 1.43$ is quite asymmetric compared to $q = 1.00$ in the full-time domain of our analysis. In this case, the pressure decrease is localised in the core of the secondary \ac{NS}, which is strongly heated by the shocks developing at the contact surface. The pressure increases in the cold external layers of the secondary \ac{NS}, which are stripped away during the dynamical evolution without being significantly heated. 

\section{Discussion}\label{sec:discussion}

Based on our results, in this section we speculate about the possible consequences for the evolution of the remnant and the emitted neutrinos. The fraction of muons is enhanced at high temperatures. Therefore, muons are expected to play a more important role in \ac{BNS} mergers characterised by larger masses and in the case of soft \acp{EOS} producing more compact \acp{NS}. Such high temperatures develop mostly at the contact interface between the two fusing cores during the first milliseconds after a merger.
In addition, \ac{CC} reactions involving $\mu^{\pm}$, $\nu_\mu$, and $\bar{\nu}_\mu$ provide a new source of opacity for muonic (anti-)neutrinos, as these reactions influence their diffusion. The neutrino surfaces of $\bar{\nu}_\mu$ and $\nu_\mu$ are expected to split in a manner similar to $\nu_e$ and $\bar{\nu}_e$ \citep{endrizzi20}, with the surface of $\bar{\nu}_\mu$ located at a larger rest mass density compared to $\nu_\mu$. Hence, the luminosity and the spectrum of the emitted $\bar{\nu}_\mu$ will deviate from that of $\nu_\mu$, with possible implications for the emitted spectrum and for the neutrino oscillations (see \cite{fischer20} for luminosity and spectra of $\nu_\mu$ and $\bar{\nu}_\mu$ in CCSN). Furthermore, the formation of trapped degenerate gases of $\nu_e$ and $\nu_\mu$ in the outer layers can favour the development of neutrino-bursts during the dynamical evolution, similar to the case of CCSN \citep{fischer20}, when dense, hot matter expands in the forming disc and becomes transparent. The change in pressure induced by muons and trapped neutrinos can change the time of the collapse of the remnant and affect the dynamical mass ejection. In particular, in (DD2, 1.00) and (SFHo, 1.00), we observed a diffuse pressure increase near the core (see discussion in \refsec{sec:pressure}), which can possibly postpone the collapse of the remnant.
For (BLh, 1.00), the drop in pressure at high density induces a softening that could speed up the collapse of more massive binaries, while the larger pressure at smaller densities could affect the low-density dynamics, including the ejecta production, the disc formation, and the stability of the remnant produced in low mass \ac{BNS} mergers.
In \refsec{sec:pressure}, we established a correlation between $P/P_{\rm sim}$ and $T_{\rm sim}$, pointing out that $P/P_{\rm sim} > 1 ~ (P/P_{\rm sim} < 1)$  when $k_{\rm B} T_{\rm sim}$ is below (above) $\sim 20$ MeV. This feature could induce positive feedback that amplifies pressure variations once muons and trapped neutrinos are included at the dynamical stage of the simulation. When the pressure decreases, the temperature of the more compressed system will increase, possibly triggering processes that lower the pressure even more. Nevertheless, when the pressure increases, the temperature of the expanded system will decrease, possibly favouring further pressure enhancement. However, feedback effects can be assessed only by including muons and trapped neutrinos in the dynamical evolution of the system. Moreover, the effect of pressure variations is, in general, non-trivial to predict. Indeed, pressure is one of the source terms in the Einstein equations and its significant increase could result in a faster collapse.

It is interesting to compare our results to the findings from \ac{CCSN} simulations that include muons \citep{bollig17, fischer20}. Our estimate of $Y_{\mu}$ exceeds the muon fraction in the proximity of the core bounce by one order of magnitude \citep{fischer20}, while it is compatible with $Y_\mu$ in the late post-bounce phase \citep{bollig17}. This is because in \ac{CCSN}e, matter density and temperature increase after the core bounce, while $Y_e$ decreases, thus approaching the regime of our analysis. Since temperatures are significantly larger in \ac{BNS} mergers than in \ac{CCSN}e, thermal processes could contribute more significantly to the creation of $\mu^{\pm}$, while they are negligible in \ac{CCSN}e \citep{bollig17, guo20, fischer20}. However, the enhancement of the net muon fraction, $Y_\mu$, with the temperature is in agreement with the results of \cite{bollig17, fischer20}.

We note that while the inclusion of muons always implies a softening of the \ac{EOS} in the post-bounce phase in \ac{CCSN}e \citep{bollig17, fischer20}, in our case, the pressure can either increase or decrease
with respect to the case where muons are neglected from the beginning. This difference is expected since in \ac{BNS} mergers, the pressure increase obtained in our analysis is driven by the muons already present in the two cold \acp{NS} (see \refsec{sec:pressure}). Notably, muons are not present in stellar iron cores before a collapse.

Next, we relate our results to state-of-the-art \ac{BNS} merger simulations that include trapped neutrinos either in post-processing \citep{perego19} or through an M1 scheme \citep[e.g.][]{foucart16a, radice22,Zappa2023}. The neutrino hierarchy resulting from our analysis is in disagreement with these works. In these previous  papers, matter is dominated by $\bar{\nu}_e$ and followed by heavy-lepton neutrinos and then $\nu_e$, whereas we found that $\bar{\nu}_\mu$ are the most abundant, followed in order by $\bar{\nu}_e$, $\overset{\scriptscriptstyle(-)}{\nu}_\tau$, $\nu_e$, and $\nu_\mu$. According to \cite{perego19}, the fluid pressure decreases after the inclusion of trapped neutrinos, but we show that it can also increase in some regions of the remnant. This is an indirect confirmation that the enhancement of pressure is mainly due to muons. Nonetheless, our estimate of the pressure decrease is compatible with the one computed in \cite{perego19}.

\section{\label{sec:conclusions}Conclusions}
In this paper, we evaluated the contribution of muons in the post-merger remnant formed during \ac{BNS} mergers as well as their impact on the trapped neutrino component during the first milliseconds after the merger. In particular, we considered a sample of numerical relativity simulations targeted at GW170817 that were performed using three different nuclear \acp{EOS} (BLh, DD2, and SFHo). Based on the remnant conditions, we estimated the abundances and the properties of muons and trapped neutrinos in post-processing, assuming weak and thermal equilibrium for rest mass densities in excess of $10^{13}{\rm g~cm^{-3}}$.
We found that a non-negligible amount of muons is present in the remnant, and they significantly alter the neutrino hierarchy, favouring the production of muon anti-neutrinos. Moreover, the new equilibrium conditions modify the pressure inside the remnant in a non-trivial way with respect to the case in which muons and trapped neutrinos are neglected.

More specifically, we found that a significant amount of muons is present at baryon densities $\rho > 10^{13} \rm{g~cm}^{-3}$ with a fraction $Y_{\mu^-} \sim 10^{-3} - 0.07$ for the BLh \ac{EOS} and $Y_{\mu^-} \sim 10^{-3} - 0.05$ for the DD2 and SFHo \acp{EOS}. The net fraction of muons is between 30$\%$ and 50-70$\%$ of the net electron fraction, with a maximum depending on the nuclear \ac{EOS}. The bulk of muons comes from the two cold \acp{NS}, while a fraction corresponding to $10^{-3} - 10^{-2}$ is created during the merger via conversion of neutrons into protons plus muons and thermal processes. Accordingly, muon production is enhanced in the high-density fusing cores and in the hot spots characterising the remnant structure during the first milliseconds after the merger. 

The presence of muons modifies the flavour hierarchy of the trapped neutrino component. In the core, matter is dominated by anti-neutrinos, with $\bar{\nu}_\mu$ being the most abundant species. Immediately outside the core and in correspondence with the hot spots, gases of trapped neutrinos and anti-neutrinos coexist, but the $\bar{\nu}_\mu$ species still dominates, followed in order by $\bar{\nu}_e$, $\overset{\scriptscriptstyle(-)}{\nu}_\tau$, $\nu_e$, and $\nu_\mu$. In contrast, in the cold matter streams at $\rho \sim 10^{13} - 10^{14} \rm{g~cm}^{-3}$, anti-neutrinos are suppressed, while neutrinos dominate and the $\nu_\mu$ species is the most abundant. Anti-neutrinos and neutrinos are degenerate in the core and in the cold matter streams, respectively. The level of degeneracy depends on the thermodynamical conditions of the remnant and on $\mu_n - \mu_p$. In particular, typical values of the degeneracy parameters of $\bar{\nu}_e$ and $\bar{\nu}_\mu$ are $\eta_{\bar{\nu}_{e,\mu}} \sim 2 - 4$, with larger (smaller) values associated with larger (smaller) $\mu_n - \mu_p$.  
Therefore, the properties of trapped (anti-)neutrinos in \ac{BNS} mergers have a non-trivial dependence on the properties of the \ac{EOS} at densities well above $n_0$. 

The presence of muons and trapped neutrinos affect the remnant pressure, whose value can decrease by $7\%$ or increase by up to $5\%$, depending on the nuclear \ac{EOS}, compared with the case in which they are neglected. The pressure drops mainly in the hot spots, where the processes driving muon production reduce the temperature and number of neutrons. The greater pressure mostly observed at low temperatures and high density is due to the muons coming from the cold \acp{NS}. 
This new result shows how adding a dof in the system microphysics does not always imply a decrease in pressure, as we could naively expect. Finally, we demonstrated that in asymmetric binaries, the pressure decrease is localised in the shocked core of the secondary \ac{NS}, while the pressure increase is found in the cold outer layers of the secondary \ac{NS}, which are stripped away by the primary \ac{NS}.

Since the procedure to post-process the simulation data is not unique, in \refapp{sec: alternative choice} we discuss the robustness of our results with respect to the arbitrary choices in our procedure. For example, we  show that when neglecting muons inside the cold \acp{NS} before merger, the major results are the same as when the presence of muons in the inspiraling \acp{NS} is considered. The comparison with the first post-processing procedure emphasises that a significant muon contribution originates from the muons already present in the cold \acp{NS}.

Our work is nevertheless limited by the post-processing approach. To understand how muons really affect the merger and post-merger dynamics, it is necessary to implement a transport scheme that includes muons as an independent dof. Also, adding the rates of the reactions involved in muon production (see \refsec{sec:analytical_estimates}) would be important for estimating the effect of muons on the neutrino opacity and luminosity, as done in CCSN simulations \citep{guo20, fischer20}. Additionally, the role of pions coupling to muons and muonic (anti-)neutrinos should be considered while computing the mean free path of low energy $\nu_\mu$ and $\bar{\nu}_\mu$ \citep{fore20}.
We also highlight that in the description of \ac{NS} matter, we have neglected the possible formation of hyperons \citep{PhysRevC.85.055806, fortin_oertel_providencia_2018} and/or that a deconfinement phase transition to quark matter may take place \citep{Weissenborn:2011qu, Klahn:2013kga, Chatterjee:2015pua, Bombaci:2016xuj, Logoteta:2019utx, Logoteta:2021iuy}. The inclusion of these additional dof will be the focus of future investigations.

This work shows, for the first time, that muons have a discernible impact on the trapped neutrino component and possibly on the thermodynamics and dynamics of \ac{BNS} mergers. Our results emphasise the need of including muons in future simulation modelling of the post-merger remnant of \ac{BNS} mergers.

\begin{acknowledgements}
The authors thank David Radice, Sebastiano Bernuzzi, Micaela Oertel, and Francesco Pederiva for useful discussions. They also thank the anonymous Referee for their valuable comments and suggestions. They acknowledge the INFN for the usage of computing and storage resources through the \texttt{tullio} cluster in Turin, and the Computational Relativity (CoRe) collaboration for providing access to the simulation data used in the work.
MB and EL acknowledge financial support from MIUR (PRIN 2020 grant 2020KB33TP). 
MB acknowledges financial support from MIUR (PRIN 2017 grant 20179ZF5KS).
\end{acknowledgements}

\bibliographystyle{aa}
\bibliography{biblio}


\begin{appendix}

\section{\label{sec: alternative choice}An alternative assumption on $Y_{l,\mu}$}

As discussed in \refsec{sec:post_proc}, there is not a unique way to assign $Y_{l,\mu}$, $Y_{l,e}$, and $e_{\rm tot}$ appearing on the left-hand side of \refeq{eq:post_proc_sys} given the outcome of the simulations presented in \refsec{sec:simulation_sample}. Therefore, in this Appendix, we consider a limiting case opposite to the one discussed in \refsec{sec:post_proc} as a different possible choice. In particular, we assume that muons are negligible in the initial conditions, that is, in the two cold NSs, and we impose $Y_{l,\mu} = 0$. Then, we identify the simulation electron fraction, $Y_e^{\text{sim}}$, and energy density, $e_{\rm sim}$, with $Y_{l,e}$ and $e_{\rm tot}$, respectively:
\begin{equation}
        Y_{l,e} = Y_e^{\text{sim}}  \qquad e_{\rm tot} = e_{\rm sim} \, 
,\end{equation}
(see from \refeq{eq:assign_yl1} to \refeq{eq:assign_energy2}).
This choice, despite being inconsistent with what we expect, has the advantage of being fully coherent with the simulation assumptions, and it provides an easier initialisation of our post-processing calculation.
In the following paragraphs, we discuss the results obtained with this choice, and we compare them with the ones discussed in the previous sections. From this point on, we refer to this particular choice as $Y_{l,\mu} = 0$ and to the one discussed in \refsec{sec:post_proc} as $Y_{l,\mu} = \tilde{Y}_{\mu}$.

\subsection{The fraction of muons}
If $Y_{l,\mu} = 0$, all the muons in the final state are produced during the merger. For all the simulations in \reftab{tab:sim_sample}, the equilibrium muon fraction at density $\rho > 10^{14} \text{g~cm}^{-3}$ is $Y_{\mu^-} \sim 10^{-3} - 3 \times 10^{-2}$. At a lower density, $\rho \sim 10^{13} - 10^{14} \text{g~cm}^{-3}$, the muon fraction is $Y_{\mu^-} \sim 10^{-3}$ if $T_{\rm sim} \gtrsim 10$ MeV; otherwise, muons are not present. We note that the amount of muons produced only during the merger is slightly smaller when $Y_{l,\mu} = \tilde{Y}_\mu$ (see \refsec{sec:muons_remnant}) because of the Pauli blocking coming from the muons already present in the cold \acp{NS}.

For all the simulations, $Y_\mu / Y_e \sim 0.3 - 0.4$, but the maximum $Y_{\mu}/Y_{e}$ is larger for DD2 and SFHo than for BLh because the latter exhibits a larger $Y_e$ (see also \reffig{fig:2}). The creation of muons is enhanced in correspondence to the hot spots for all the \acp{EOS}. In the case of (BLh, 1.00), muon production also increases in the density region $\rho > 10^{15} \text{g~cm}^{-3}$. This effect depends on the significantly larger electron fraction obtained with the BLh \ac{EOS} since the more degenerate electrons block the conversion of neutrons into protons plus electrons, favouring instead the conversion into protons plus muons.

\subsection{The trapping of neutrinos}
The initial assumption on $Y_{l,\mu}$ does not significantly affect the behaviour of electron (anti-)neutrinos, but it does imply discernible consequences on the muon neutrinos. If $Y_{l,\mu} = 0$, muon neutrinos appear only in correspondence to the hot spots and the warm matter streams, with a maximum fraction $Y_{\nu_\mu} \sim 0.01$. All $\nu_\mu$ are produced together with $\bar{\nu}_\mu$ since there is not an initial muon fraction that can be converted into $\nu_\mu$. In contrast, the amount of $\bar{\nu}_\mu$ at equilibrium is strongly enhanced to guarantee the conservation of $Y_{l,\mu}$. Therefore, $Y_{\bar{\nu}_\mu} \approx Y_{\mu^-}$ for all the simulations considered,
and $\bar{\nu}_\mu$ constitute the most abundant species amongst all flavour (anti-)neutrinos in the regions that are not dominated by thermal production.

The differences in the equilibrium fractions of neutrinos depending on the chosen $Y_{l,\mu}$ reflect the differences in the degeneracy parameters.
For all the simulations $\eta_{\nu_\mu} \leq 0$ and the production of $\nu_\mu,$  only thermal and trapped degenerate $\nu_\mu$ are suppressed. In addition, the gases of $\bar{\nu}_e$ and $\bar{\nu}_\mu$ trapped in the remnant core are more degenerate when $Y_{l,\mu} = 0$ compared to when $Y_{l,\mu} = \tilde{Y}_{\mu}$. This follows in part from the enhanced production of $\bar{\nu}_\mu$ when $Y_{l,\mu} = 0$, but it also depends on the equilibrium temperature $T$, which is systematically larger when $Y_{l,\mu} = \tilde{Y}_{\mu}$. Actually, in the latter case, the system lowers the fraction of electrons and neutrons to take into account the fraction of muons from the cold \acp{NS}, which practically means reducing the system degeneracy by increasing the temperature (see the discussions in \refsec{sec:post_proc} and \refsec{sec:pressure}).

We stress that the presence of a trapped degenerate component of $\bar{\nu}_e$ and $\bar{\nu}_\mu$ in the core and of $\nu_e$ in the outer layers of the remnant is a robust prediction independent of the initial assumption on $Y_{l,\mu}$. Instead, the presence of a degenerate gas of $\nu_\mu$ in the cold matter streams depends on the initial choice of $Y_{l,\mu}$.

\subsection{The changes in pressure}
In conclusion, we analysed the ratio $P/P_{\rm sim}$ where $P$ is computed in post-processing under the assumption $Y_{l,\mu} = 0$. In this case, we did not observe a significant pressure increase after muons were introduced (i.e. $P/P_{\rm sim} > 1$) for any of the simulations in \reftab{tab:sim_sample}. This result is expected since, within this assumption, we neglected the muons from the cold \acp{NS}, which are indeed responsible for the pressure increase when $Y_{l, \mu} = \tilde{Y}_{\mu}$, as discussed in \refsec{sec:pressure}. In fact, the amount and the spatial distribution of the pressure decrease are common to all the simulations irrespective of the initial assumption on $Y_{l, \mu}$. In particular, the pressure decrease is still favoured in correspondence to the hot spots because, in these regions, the final equilibrium fractions of muons and (anti-)neutrinos are comparable for the two choices of $Y_{l,\mu}$.

\end{appendix}

\end{document}